\documentclass[sigplan, 10pt, pbalance, screen, nonacm]{acmart}
\acmArticleType{Research}

\acmConference{sumbitted to ArXiv}
\acmYear{}
\acmISBN{}        
\acmDOI{}         
\startPage{1}

\setcopyright{none}

\usepackage{etex}               
\usepackage{amscd}              
\usepackage{amsmath}            
\usepackage{stmaryrd}           
\usepackage{mathrsfs}           
\usepackage{times}              
\usepackage{latexsym}           
\usepackage{hhline}             
\usepackage{multicol}           
\usepackage{multirow}           
\usepackage{rotating}           
\usepackage[latin1]{inputenc}   
\usepackage{calc}               
\usepackage{graphicx}           
\usepackage{ifthen}             
\usepackage{colortbl}           
\usepackage{booktabs}           
\usepackage{array}              
\usepackage{booktabs}           
\usepackage{enumitem}           
\usepackage{leftidx}            
\usepackage{longtable}          
\usepackage{mathtools}          
\usepackage{framed}             
\usepackage{setspace}           
\usepackage{srcltx}             
\usepackage{epstopdf}           
\usepackage{todonotes}			
\usepackage{fancybox}           
\usepackage{nicematrix}         

\usepackage[algo2e, plain, vlined]{algorithm2e}

\usepackage[indent=7pt, skip=1pt]{parskip}

\usepackage{pst-all}            
\usepackage{pst-poly}           
\usepackage{pst-node}			
\usepackage{pstricks}			
\usepackage{multido}            
\usepackage{pst-2dplot}

\usepackage{tikz}
\usetikzlibrary{shapes}
\usetikzlibrary{arrows}
\usetikzlibrary{trees}
\usetikzlibrary{matrix}
\usetikzlibrary{automata}
\usetikzlibrary{plotmarks}
\usetikzlibrary{external}

\usepackage{caption}
\usepackage{subcaption}
\usepackage{extarrows}

\newcommand{\pspar}{\psset{arrows=->, arrowscale=1.3, linewidth=1pt, linecolor=black, linestyle=solid, rowsep=1.5cm, colsep=1.5cm, levelsep=1.5cm, nodesep=3pt, labelsep=3pt, arcangle=30, border=0.05cm}}

\newcommand{\autps}{\pspar}

\newenvironment{automaton}[1] {
\tabcolsep=0.0cm\def\arraystretch{0}
\begin{tabular}[c]{c}
\autps\psset{#1}
\begin{psmatrix}
} {
\end{psmatrix}
\end{tabular}
}

\newcommand{\treeps}{\psset{arrows=-, arrowscale=2, treefit=loose, treesep=0.5cm, linewidth=1pt,
linecolor=black, linestyle=solid, levelsep=1.5cm, nodesep=4pt, labelsep=0.25cm}}

\newenvironment{syntaxtree}[1] {
\tabcolsep=0.0cm\def\arraystretch{0}
\begin{tabular}[c]{c}
\treeps\psset{#1}
} {
\end{tabular}
}

\def\ZenodoHREF{\href{https://zenodo.org/records/14219357}{Zenodo}}

\def\REgenHREF{\href{https://github.com/FLC-project/REgen}{REgen}}

\def\fasta{\url{https://open.oregonstate.education/computationalbiology/chapter/patterns-regular-expressions}}

\def\fastaHREF{\href{https://open.oregonstate.education/computationalbiology/chapter/patterns-regular-expressions}{fasta}}

\def\traffic{\url{https://zenodo.org/record/5789064\#.ZCHHQ9LP0eM}}

\def\trafficHREF{\href{https://zenodo.org/record/5789064\#.ZCHHQ9LP0eM}{traffic}}

\def\bible{\url{https://www.gliscritti.it/dchiesa/bibbia_cei08/indice.htm}}

\def\bibleHREF{\href{https://www.gliscritti.it/dchiesa/bibbia_cei08/indice.htm}{bible}}

\def\Ondrik{\url{https://github.com/ondrik/automata-benchmarks?tab=readme-ov-file}}

\def\OndrikHREF{\href{https://github.com/ondrik/automata-benchmarks?tab=readme-ov-file}{Ondrik}}

\def\SNORTHREF{\href{http://www.snort.org}{SNORT}}

\newcommand{\set}[1]{\left\{ #1 \right\}}

\newcommand{\asgn}{\vcentcolon =}

\newcommand {\interfun} {\emph{if}}

\begin{document}

\title[Minimizing speculation overhead in a parallel recognizer for regular texts]{Minimizing speculation overhead \\ in a parallel recognizer for regular texts}


\author{Angelo Borsotti}
\orcid{0000-0001-6444-0361}              
\affiliation{
  \institution{Politecnico di Milano}    
  \streetaddress{Piazza Leonardo Da Vinci $32$}
  \city{Milano}
  \postcode{$20133$}
  \country{Italy}                        
}
\email{angelo.borsotti.@mail.polimi.it}  

\author{Luca Breveglieri}
\orcid{0000-0001-5294-6840}              
\affiliation{
  \institution{Politecnico di Milano}    
  \streetaddress{Piazza Leonardo Da Vinci $32$}
  \city{Milano}
  \postcode{$20133$}
  \country{Italy}                        
}
\email{luca.breveglieri@polimi.it}       

\author{Stefano Crespi Reghizzi}
\orcid{0000-0001-5061-7402}              
\affiliation{
  \institution{Politecnico di Milano, CNR-IEIIT}    
  \streetaddress{Piazza Leonardo Da Vinci $32$}
  \city{Milano}
  \postcode{$20133$}
  \country{Italy}                        
}
\email{stefano.crespireghizzi@polimi.it}        

\author{Angelo Morzenti}
\orcid{0000-0002-6469-2929}              
\affiliation{
  \institution{Politecnico di Milano}    
  \streetaddress{Piazza Leonardo Da Vinci $32$}
  \city{Milano}
  \postcode{$20133$}
  \country{Italy}                        
}
\email{angelo.morzenti@polimi.it}        

\begin{abstract}
Speculative data-parallel algorithms for language recognition have been widely experimented for various types of finite-state automata (FA), deterministic (DFA) and nondeterministic (NFA), often derived from  regular expressions (RE). Such an algorithm  cuts the input string into chunks,   independently recognizes each chunk in parallel by  means of identical FAs, and at last joins the  chunk results and checks  overall consistency.  In chunk recognition, it is necessary to speculatively start the FAs in any state, thus causing an overhead that  reduces the speedup compared to a serial algorithm. Existing data-parallel DFA-based recognizers suffer from the excessive number of starting states, and the NFA-based ones suffer from the  number of nondeterministic transitions.
\par
Our data-parallel algorithm is based on the new FA type called  reduced interface DFA (RI-DFA), which  minimizes the speculation overhead  without incurring in the penalty of nondeterministic transitions or of  impractically enlarged DFA machines. The  algorithm is proved to be correct and theoretically efficient, because it combines the state-reduction of an NFA  with the speed of deterministic transitions, thus improving on  both DFA-based and NFA-based existing implementations. The practical applicability of the RI-DFA approach is confirmed by a quantitative comparison of the number of starting states for a large public benchmark of complex FAs. On multi-core computing architectures, the RI-DFA recognizer is much faster than the  NFA-based one on all benchmarks, while it matches  the DFA-based one on some benchmarks and performs much better on some others. The extra time cost needed to construct an RI-DFA compared to a DFA is moderate and is compatible with a practical use.
\end{abstract}


\maketitle

\section{Introduction} \label{sec:introduction}
The recognition of regular languages by a finite-state automaton (FA) is one of the most widely used basic algorithms, and has been extensively investigated to take advantage of parallel computing architectures. Parallel recognition algorithms exist in many versions optimized for different  architectures, such as reconfigurable hardware (FPGA), SIMD and GPU accelerators, vectorial machines and multi-processor / multi-core machines. Here we present a new  data-parallel  algorithm based on a novel type of multi-entry DFA (a deterministic FA with multiple initial states)  that is efficient in theory and performs well on commodity multi-core computers.
\par\noindent
\textbf{\emph{Theoretical challenge}.} The baseline is the classical speculative data-parallel algorithm (for short CSDPA), see, e.g., \cite{DBLP:journals/cacm/HillisS86} for an early reference, \cite{DBLP:conf/asplos/MytkowiczMS14} for theoretical and practical aspects, and \cite{DBLP:conf/wia/HolubS09} for code description. After splitting the input text into a number $c \geq 1$ of chunks, CSDPA operates in two phases. First, the \emph{reach} phase  processes in parallel each chunk by using  the same \emph{chunk automaton} (CA), and answers two questions: is the chunk a legal substring for the input language, and which pairs of starting and ending FA states are involved in substring recognition. Then, the \emph{join} phase  checks  the consistency of the sets of state pairs, for any two adjacent chunks. The  hot-spot is in the reach phase, which depends on the input length. The  number $\vert \, Q \, \vert$ of CA states  affects the recognition time, because a CA (except for the first one) does not know the last state that was reached by the upstream  CA, and  it is forced to  start  a speculative run in each state. Thus in the worst case, the number of transitions needed to parse a string of length $n \geq 1$, segmented into $c$ chunks, is $\mathcal O \, (n \times \vert \, Q \, \vert)$, instead of just $\mathcal O \, (n)$ as  in the serial recognition. Since the size $\vert \, Q \, \vert$ of the CA is  often much larger than the number of available computing cores, a parallel recognizer  may turn out to be  slower than a serial one! The join phase is lighter and is typically serial, since it does not depend on $n$ but only on $c$, with  $c \ll n$.
\par
We briefly discuss the state-of-the art of some CSDPA variants. A DFA  may incur in a very large speculation overhead. Researchers have also considered an NFA, since the number of states $\vert \, Q_N \, \vert$ can be made less than or equal to $\vert \, Q \, \vert$. The cardinality $\vert \, Q \, \vert$ of the minimal equivalent DFA is $\mathcal O \, \left( 2^{\vert \, Q_N \, \vert} \right)$ in the worst case \cite{DBLP:conf/focs/MeyerF71}. Unfortunately, NFA simulation is expensive: for instance the time complexity of the parallel algorithm \cite{DBLP:journals/jacm/LadnerF80}  based on prefix-sum is $\mathcal O \, \big( n^3 \times \log m \big)$, with $m$ processors. In practice, the state reduction brought by an NFA does not pay in general. This is a finding  of prior experimental studies on CSDPA using NFA, e.g.,~\cite{FuZheLi2017}, and is  also confirmed by our measurements (see Tab.~\ref{tab:maxspeedup&trans}).  However, an NFA  may offer advantages for some specific applications. For instance, it has been used for  packet content scanning in the \SNORTHREF~intrusion detection system. Others, e.g.,~\cite{YangEtAl_semiDFA2011}, have proposed to  pre-process the DFA in order to select as initial the states that are likely to  be successful over the input texts considered; the remaining states are left for serial execution in case of failure. We did not consider such an approach, as it  depends on  language statistics, while we aim at a good performance in general.
\par
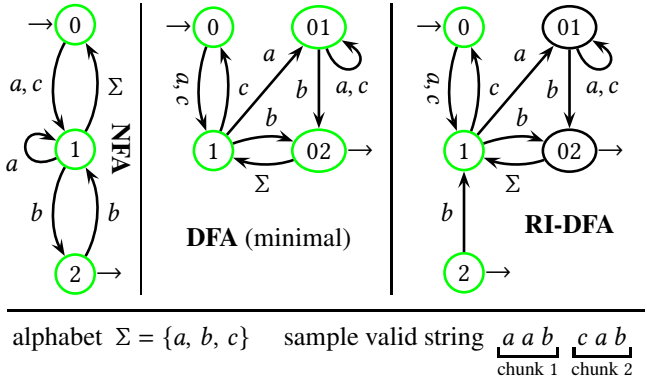
\begin{figure}[b]
\begin{center}
\vspace{0.25cm}
\tabcolsep=0.5cm
\begin{tabular}{@{\hspace{0.0cm}}c|c|c@{\hspace{0.0cm}}}
\begin{automaton}{nodesep=0pt, rowsep=1.1cm, colsep=0.75cm, border=0.0cm}
\circlenode[linecolor=green] 0 {$0$} \\

\circlenode[linecolor=green] 1 {$1$} \\

\circlenode[linecolor=green] 2 {$2$}

\nput[labelsep=0pt] {180} 0 {$\rightarrow$}

\nput[labelsep=0pt] {0} 2 {$\rightarrow$}

\nput[labelsep=6pt, rot=-90] {0} 1 {\bf NFA}

\ncarc[arcangle=-30] 0 1 \nbput{$a, c$}

\nccurve[angleA=-155, angleB=155, ncurv=5] 1 1 \naput[npos=0.375]{$a$}

\ncarc[arcangle=-30] 1 2 \nbput{$b$}

\ncarc[arcangle=-30] 2 1 \nbput{$b$}

\ncarc[arcangle=-30] 1 0 \nbput{$\Sigma$}
\end{automaton}
&
\begin{automaton}{nodesep=0pt, rowsep=1.05cm, colsep=0.75cm, border=0.0cm}
\circlenode[linecolor=green] 0 {$0$} & \ovalnode[linecolor=green] {01} {$01$} \\

\circlenode[linecolor=green] 1 {$1$} & \ovalnode[linecolor=green] {02} {$02$} \\

\phantom{\circlenode 2 {$2$}}

\nput[labelsep=0pt] {180} 0 {$\rightarrow$}

\nput[labelsep=0pt] {0} {02} {$\rightarrow$}

\nput[labelsep=0.75] {-90} 1 {\hspace{1.375cm} {\bf DFA} (minimal)}

\ncarc[arcangle=-20] 0 1 \nbput[nrot=:U]{$a, c$}

\ncarc[arcangle=-20] 1 0 \nbput{$c$}

\ncline 1 {01} \naput[npos=0.75]{$a$}

\ncarc[arcangle=20] 1 {02} \naput[npos=0.625]{$b$}

\ncarc[arcangle=20] {02} 1 \naput[npos=0.5]{$\Sigma$}

\nccurve[angleA=-70, angleB=-30, ncurv=5] {01} {01} \nbput[npos=0.4]{$a, c$}

\ncline {01} {02} \nbput{$b$}
\end{automaton}
&
\begin{automaton}{nodesep=0pt, rowsep=1.05cm, colsep=0.75cm, border=0.0cm}
\circlenode[linecolor=green] 0 {$0$} & \ovalnode {01} {$01$} \\

\circlenode[linecolor=green] 1 {$1$} & \ovalnode {02} {$02$} \\

\circlenode[linecolor=green] 2 {$2$}

\nput[labelsep=0pt] {180} 0 {$\rightarrow$}

\nput[labelsep=0.55cm] {-90} {02} {\bf RI-DFA}

\nput[labelsep=0pt] {0} 2 {$\rightarrow$}

\nput[labelsep=0pt] {0} {02} {$\rightarrow$}

\ncarc[arcangle=-20] 0 1 \nbput[nrot=:U]{$a, c$}

\ncarc[arcangle=-20] 1 0 \nbput{$c$}

\ncline 1 {01} \naput[npos=0.75]{$a$}

\ncarc[arcangle=20] 1 {02} \naput[npos=0.625]{$b$}

\ncarc[arcangle=20] {02} 1 \naput[npos=0.5]{$\Sigma$}

\nccurve[angleA=-70, angleB=-30, ncurv=5] {01} {01} \nbput[npos=0.4]{$a, c$}

\ncline {01} {02} \nbput{$b$}

\ncline 2 1 \naput{$b$}
\end{automaton}
\end{tabular}
\vspace{0.2cm}
\hrule
\vspace{0.2cm}
alphabet \ $\Sigma = \set{a, \, b, \, c}$ \quad
sample valid string \ $\underbracket[0.75pt]{ \, a \ a \ b \, }_\text{chunk $1$}$ \ $\underbracket[0.75pt]{ \, c \ a \ b \, }_\text{chunk $2$}$
\vspace{0.2cm}
\hrule
\vspace{0.2cm}
\tabcolsep=0.0cm
\arraycolsep=0.0cm
\begin{tabular}[c]{m{1.5cm}@{\hspace{0.4cm}}llc}
& \multicolumn{3}{c}{\emph{runs of the {\rm CAs} and their transition counting}} \\ \cline{2-4}
\emph{used {\rm CA}} & \emph{chunk $1$} & \emph{chunk $2$} & \emph{total} \\ \toprule
min DFA  \par\noindent\scriptsize classic \par method & \, \scriptsize $0 \xrightarrow a 1 \xrightarrow a 01 \xrightarrow b \rnode{DFA1}{02}$ & \, \scriptsize $\begin{array}{l} 0 \xrightarrow c 1 \xrightarrow a 01 \xrightarrow b 02 \\ 1 \xrightarrow c 0 \xrightarrow a 1 \xrightarrow b 02 \\ 01 \xrightarrow c 01 \xrightarrow a 01 \xrightarrow b 02 \\ \rnode{DFA2}{02} \xrightarrow c 1 \xrightarrow a 01 \xrightarrow b 02 \end{array}$ & $15$ \nccurve[arrows=-, linewidth=0.5pt, linestyle=dashed, linecolor=red, angleA=-15, angleB=165, nodesep=3pt, ncurv=1] {DFA1} {DFA2} \naput[labelsep=2pt, npos=0.25, nrot=:U] {\red\tiny join} \\ \midrule
NFA \par \scriptsize classic \par optimized \par method & \scriptsize
\tabcolsep=0.0cm
\def\arraystretch{0}
\begin{tabular}{c}
\begin{syntaxtree}{labelsep=2.5pt, nodesep=1.5pt, arrows=->, arrowscale=1, levelsep=0.625cm, linewidth=0.5pt, treemode=R, treesep=0.25cm}
\pstree{\TR{$0$}} {
    \pstree{\TR{$1$}} {\taput{$a$}
        \pstree{\TR{$0$}} {\taput[tpos=0.3]{$a$}}
        \pstree[treesep=0.25cm]{\TR{$1$}} {\tbput[tpos=0.3]{$a$}
            \pstree{\TR[name=NFA1]{$0$}} {\taput[tpos=0.4]{$b$}}
            \pstree{\TR{$2$}} {\tbput[tpos=0.4]{$b$}}
        }
    }
}
\end{syntaxtree}
\end{tabular}
& \scriptsize
\tabcolsep=0.0cm
\begin{tabular}{l}
\begin{tabular}{c}
\begin{syntaxtree}{labelsep=2.5pt, nodesep=1.5pt, arrows=->, arrowscale=1, levelsep=0.625cm, linewidth=0.5pt, treemode=R, treesep=0.25cm}
\pstree{\TR[name=NFA2]{$0$}} {
    \pstree{\TR{$1$}} {\taput{$c$}
        \pstree{\TR{$0$}} {\taput[tpos=0.3]{$a$}}
        \pstree[treesep=0.25cm]{\TR{$1$}} {\tbput[tpos=0.3]{$a$}
            \pstree{\TR{$0$}} {\taput[tpos=0.4]{$b$}}
            \pstree{\TR{$2$}} {\tbput[tpos=0.4]{$b$}}
        }
    }
}
\end{syntaxtree}
\end{tabular}
\\[0.75cm]
\begin{tabular}{c}
\begin{syntaxtree}{labelsep=2.5pt, nodesep=1.5pt, arrows=->, arrowscale=1, levelsep=0.625cm, linewidth=0.5pt, treemode=R, treesep=0.25cm}
\pstree{\TR{$1$}} {
    \pstree{\TR{$0$}} {\taput{$c$}
        \pstree[treesep=0.25cm]{\TR{$1$}} {\taput[tpos=0.3]{$a$}
            \pstree{\TR{$0$}} {\taput[tpos=0.4]{$b$}}
            \pstree{\TR{$2$}} {\tbput[tpos=0.4]{$b$}}
        }
    }
}
\end{syntaxtree}
\end{tabular}
\end{tabular}
& $14$ \nccurve[arrows=-, linewidth=0.5pt, linestyle=dashed, linecolor=red, angleA=15, angleB=-165, nodesep=3pt, ncurv=1] {NFA1} {NFA2} \naput[labelsep=2pt, npos=0.75, nrot=:U] {\red\tiny join} \\
[1.0cm] \midrule
RI-DFA \par \scriptsize new method & \, \scriptsize $0 \xrightarrow a 1 \xrightarrow a 01 \xrightarrow b \rnode{RIDFA1}{0}2$  & \, \scriptsize $\begin{array}{l} \rnode{RIDFA2}{0} \xrightarrow c 1 \xrightarrow a 01 \xrightarrow b 02 \\ 1 \xrightarrow c 0 \xrightarrow a 1 \xrightarrow b 02 \end{array}$ & $9$ \nccurve[arrows=-, linewidth=0.5pt, linestyle=dashed, linecolor=red, angleA=-45, angleB=-165, nodesep=3pt, ncurvA=0.75, ncurvB=1] {RIDFA1} {RIDFA2} \nbput[labelsep=1pt, npos=0.175] {\red\tiny $02$ includes $0$} \naput[labelsep=2pt, npos=0.75, nrot=:U] {\red\tiny join} \\
\end{tabular}
\vspace{0.25cm}
\end{center}
\caption{Top: NFA with the equivalent powerset DFA (minimal) and the new RI-DFA, over the alphabet $\Sigma$. The states that act as initial in the CA are in green. Bottom:  transitions executed by the reach phase for the string ``$aabcab$'' divided in two chunks, join of the ending and starting states of adjacent chunks, and number of transitions.} \label{fig:RIDintroductory}
\end{figure}
So far we have focused on the number of CA starting states, but  also  the whole state cardinality $\vert \, Q \, \vert$   may negatively impact on performance for very large CAs, which can cause too many cache misses \cite{DBLP:conf/asplos/MytkowiczMS14}. For this, the use of NFA may be convenient also for a different reason: CA construction requires $\text{RE} \to \text{DFA}$ or $\text{NFA} \to \text{DFA}$ conversion algorithms, which have exponential time-complexity in $\vert \, Q \, \vert$ and may be too slow for on-line use. In most other cases, it is fair to assume that the state transition cost is independent of CA size.
\par
Eventually, we mention the Simultaneous Finite Automata (SFA)  algorithm, which completely avoids speculation at the cost of state  explosion \cite{DBLP:conf/icpp/SinyaMS13}; it is assessed in~\cite{FuZheLi2017}. Given a deterministic CA  where all the states are initial, the equivalent  SFA is a much larger  DFA, each state of which is characterized by a set of pairs $(i, \, f)$, where $i$ and $f$ are respectively the starting and arrival  states of  a run on the  CA. Therefore,  speculation disappears since the multiple parallel runs of the CA are mapped on the  single run between two SFA states characterized by the corresponding sets of pairs $(i, \, f)$. The drawback is an explosion of the number of states, with the consequence that the  SFA construction takes too long and recognition may suffer from cache misses. For an RE  of moderate size,  the construction can be thousand times slower than  for a DFA \cite{FuZheLi2017}. Later research \cite{DBLP:conf/icpp/JungPBB17} is striving to reduce the construction time and state-transition costs.
\par
\textbf{\emph{Sum up}}: \emph{for data-parallel recognition on multi-core computers,  the challenge is to curb the speculation overhead by reducing the number of initial states of the chunk automaton.}
\par\noindent
\textbf{\emph{Theoretical contribution}.} We introduce a novel type of deterministic CA, called \emph{reduced-interface} DFA (RI-DFA): it reduces the CA interface size down to the NFA size $\vert \, Q_N \, \vert$ and preserves determinism. The RI-DFA is a DFA with multiple initial states, a model  considered in some theoretical papers. The earliest one,  called \emph{multi-entry finite automaton} (MEFA) \cite{DBLP:journals/jcss/GillK74}, requires all states to be initial. It was later renamed MDFA, allowing the  initial states to be a proper subset. Thus,  RI-DFA is a special type of MDFA with new features targeted to parallel recognizers. To our knowledge, such  a device  is unknown in the theory (a series of theoretical studies on MDFAs is \cite{DBLP:journals/jcss/GillK74,DBLP:journals/jcss/GalilS76,DBLP:journals/jcss/VelosoG79, DBLP:journals/jalc/Kappes00, DBLP:journals/jalc/HolzerSY01, DBLP:conf/dcfs/PalioudakisSA13a}).
\par
We anticipate the construction of RI-DFA (Sect.~\ref{sec:reduced-interface-device}) in Fig.~\ref{fig:RIDintroductory}. The construction starting from an  NFA has the same complexity as that of a DFA. The NFA, minimal DFA and RI-DFA shown in Fig.~\ref{fig:RIDintroductory} are equivalent. We compare their performance as CAs. All the states $\set{0, 1, 2}$ of the NFA and $\set{0, 1, 01, 02}$ of the DFA must be taken as  initial. As expected, the NFA has fewer states: $3 = \vert \, Q_N \, \vert < \vert \, Q \, \vert = 4$ for DFA. The RI-DFA CA has five states, $Q_\text{RI-DFA} = \set{0, 1, 2, 01, 02}$, but only the states $\set{0, 1, 2}$ are initial, i.e., exactly those in $Q_N$. The graph of RI-DFA may look as the superimposition of the NFA and DFA  graphs,  but its definition is more subtle. Intuitively, it is obtained  by starting in any  NFA  state, namely $0$, $1$ or $2$, and then jumping into the existing DFA states. Since states $0$ and $1$ are already present in  DFA, their subgraphs in NFA and in RI-DFA are identical, and it remains  to explain what happens starting in state $2$. Since there exists the transition $2 \xrightarrow[\text{NFA}] b 1$, we draw the edge $2 \xrightarrow[\text{RI-DFA}] b 1$. The  remaining DFA states $01$ and $02$ are not initial for the CA, because they do not belong to the NFA. Thus,  RI-DFA has fewer initial states than DFA and  may perform fewer speculative transitions. The  count of transitions for the  cases DFA, NFA and RI-DFA  is in Fig.~\ref{fig:RIDintroductory}, bottom, for the sample string $aabcab$, segmented in $c = 2$  chunks. The totals are  $15$, $14$ and $9$ for the  DFA, NFA and RI-DFA, respectively. Such numbers  approximately  measure  the overall work done by the recognizer. The serial DFA recognizer  executes exactly $n =\vert \, aabcab \, \vert = 6$ transitions on the whole input, therefore the exceeding transitions in Fig.~\ref{fig:RIDintroductory} measure the extra work due to speculation, which is thus minimal when using the RI-DFA as CA. In Sect.~\ref{sec:experimental-results} such a  finding is confirmed and is quantified for  representative benchmarks.
\par
Of course, the best saving in transition counts for RI-DFA would be obtained  when the NFA is state-wise minimal.  Unfortunately, no polynomial-time algorithms are known for NFA minimization \cite{DBLP:conf/focs/MeyerS72, DBLP:journals/siamcomp/JiangR93}. Therefore, we could not rely on   minimal NFAs  in our experiments, yet we obtained significant speedup thanks to our second  theoretical advance. It partially compensates for the lack of NFA minimality,
by means of a novel method for reducing RI-DFA initial states, by carefully exploiting the classic  state-equivalence (a.k.a.  undistinguishability) relation of DFAs. Notice that the  minimization of DFA states  does not apply  to an RI-DFA, since merging undistinguishable states would produce a machine unsuitable as a CA. A careful application of the minimization algorithm is however possible (see Sect.~\ref{subsec:furtherReduction}), and in many cases the RI-DFA achieves a significantly better performance than the  CSDPA based on DFA or NFA.
\par
Furthermore, it is important to say that the optimization based on RI-DFA is compatible with most other optimizations that have been proposed for finite-state machines, and it would be interesting to experiment this optimization in  combination with the existing ones.
\par
At last, there may be  some concern that   the higher complexity of the $\text{NFA} \to \text{RI-DFA}$ construction over the classical $\text{NFA} \to \text{DFA}$ transformation might penalize practical use. This is not the case, as we have found that for a large public collection of big NFAs the construction time is moderately higher and remains very acceptable for practical application.
\par\noindent
\textbf{\emph{Practical contribution}.} One may wonder about the need for a new parallel algorithm. Parallel recognition
is needed when the text to recognize is large: the text is split into chunks on which recognition can
be done in parallel. For the first chunk, the initial state of a CA is known and recognition is real-time. But for any successive chunk, the CA  cannot await to know the final state(s) reached by the previous (upstream)  CA, because that would imply  serial execution. This means that all CAs, but the first one,  must start from every  state, and at the end must discard the runs that do not start from a final state reached by the upstream CA.
It is not sure that this  is too big a penalty. In practice, previous experience has evidenced that in many real-world benchmarks most CA runs
quickly terminate before reaching the chunk end. However, the fewer the CA starting states, the fewer the chunk recognitions, and the smaller the penalty caused by speculation. The classic CSDPA algorithm uses a DFA as CA, which must start in all the states. On the other hand, our \emph{reduced-interface device} (RID) algorithm uses the new type of FA (RI-DFA), which has fewer initial states, as many as the NFA. Such a CA  emulates, in a deterministic way, all the runs  an NFA would do.
\par
In Sect.~\ref{subsec:experimental-minimality} we compare  the number of initial states of the NFAs, minimal DFAs and RI-DFAs, for a large number (over $1000$) of  big automata available in the public \emph{\OndrikHREF} collection. We so obtain a first assessment of  the potential superiority of the RI-DFA technique: for over $90\%$ of the automata the number of states of  the equivalent  RI-DFA  is significantly smaller than the number of states of the equivalent minimal DFA. In Sect.~\ref{subsec:experimental-parallel-recognition} we report the relative speed of the three CSDPA variants based on NFA, DFA and RI-DFA chunk automata, on a shared-memory $64$-core parallel computer. The data sets are  three public benchmarks pertaining to string pattern matching in books, biological data, and system log files of network traffic, plus two synthetic benchmarks. We have implemented in Java language the three variants of our algorithm, available on \ZenodoHREF. The RI-DFA variant is  faster than the NFA one in all cases; it performs as the DFA variant $\pm 10\%$ on two of the public benchmarks, and on long input it dominates the DFA variant in the remaining cases.
\par\noindent
\textbf{\emph{Paper organization}.} Sect.~\ref{sec:classic-device} recalls the classic speculative data-parallel  algorithm (CSDPA) with its DFA and NFA variants. Sect.~\ref{sec:reduced-interface-device} presents the reduced-interface CA and its construction,  proves correctness, and ends with an  optimization for further reducing the initial states. Sect.~\ref{sec:experimental-results} presents and discusses experimental results. It starts with the choice of benchmarks. Then it reports the reduction of the number of initial states for the \emph{\OndrikHREF} collection of automata. Eventually it reports the  measurements of the count of state transitions and it ends with the speedup evaluation on a multi-core computer. Sect.~\ref{sec:conclusion} concludes and outlines future developments. Comparisons and references to  related work are placed by a best fit criterion in all the sections.
\section{Preliminaries on data-parallel recognition} \label{sec:classic-device}
We need to precisely describe the classic speculative algorithm CSDPA, following in particular \cite{DBLP:conf/wia/HolubS09}, by means of standard concepts of automata theory. Our description is sufficiently general to fit, not only  the classic case, but also  the new development in Sect.~\ref{sec:reduced-interface-device}. The input alphabet is $\Sigma$, the regular language is $L$, the input string is $x \in L$, of length $\vert \, x \, \vert = n \geq 1$, and a segmentation of $x$ into $c \geq 2$ chunks $y_i$ is $x = y_1 \, y_2 \, \ldots \, y_c$, with $y_i \in \Sigma^+$. We assume that language $L$ is recognized by a given finite automaton $A$, with state set $Q$, alphabet $\Sigma$, state-transition graph $\delta$, initial state $q_0$ and final state set $F$. Automaton $A$ may be nondeterministic or deterministic, depending on the case  considered.
\par
Algorithm CSDPA is abstractly modeled by a recognition  \emph{device} consisting of a series $A_1 \ldots A_c$ of $c \geq 2$  identical \emph{chunk automata} (CA), called $A_i$, which are obtained from automaton $A$. A run  starts in the initial state  $q_0$ and accepts in the final states $F$. A generic CA $A_i$ is as follows:
\[
A_i = \big( \, Q, \, \Sigma, \, \delta^A, \, I_i^A, \, F_i^A \, \big)
\]
where, as usual,  $Q$ is the state set, and $I_i^A = F_i^A = Q$ are the initial and final state sets, respectively. Notice that all states are both initial and final, with the optimization for the first CA where  $I_1^A = \set{q_0}$. Then $\delta^A = \delta$ represents the state-transition graph, identical to that of the recognizer $A$ of $L$. Depending on $\delta^A$ being a function or a binary relation, we call a CA deterministic or nondeterministic. Viewed in isolation, a CA $A_i$ is a recognizer of all the substrings of $L$, and additionally it has the  capability of recognizing whether a substring is a prefix or a suffix of $L$. A chunk  $y_i$ is  accepted by $A_i$ if it is  consumed by at least one  run in $\delta^A \, (q, \, y_i)$, for some state $q \in I_i^A$, denoted for brevity as $\delta^A \, (I_i^A, \, y_i)$.
\par
The device  operates \emph{serially} if at any time only one of the CAs $A_i$ is active, all the upstream CAs have successfully  finished, and all the downstream CAs are awaiting. In \emph{serial operation}, the CA $A_1$ processes chunk $y_1$ starting in state $I_1^A = \set{q_0}$, and passes to CA $A_2$ the set of \emph{last active  states} $\text{LAS}_1 = \delta^A \, (I_1^A, \, y_1)$ as those to start from. Similarly,  each CA $A_i$, for each $2 \leq i \leq c$, is initialized with the  states $I_i^A = \text{LAS}_{i-1}$ and passes (except for the last CA $A_c$) the set $\text{LAS}_i = \delta^A \, (I_i^A, \, y_i)$ to its next downstream CA $A_{i+1}$. The device accepts if the last set LAS contains a final state, i.e., $\text{LAS}_c \, \cap \,  F \not = \emptyset$.
\par
Here we  focus on the computational load for recognizing the input string $x$,  expressed as the total number of state transitions executed, since we may disregard the effort for checking  the  acceptance condition, which is independent of the input size. In serial operation, if the CA type is  deterministic, the overall number of transitions is  $\sum_{i = 1}^c \vert \, y_i \, \vert = \vert \, x \, \vert = n$, and is independent of language $L$ and of the CA size. On the other hand, if the CA type is nondeterministic,  the number of transitions may exceed the length $n$ and  depends on the degree of nondeterminism of the CA, as well as on the input.
\par
In \emph{parallel operation}, all CAs start in parallel in the initial states $I_i^A = Q$, again with the optimization of $I_1^A$ to $\set{q_0}$, and  proceed to recognize the chunks. Each CA $A_i$, for each $1 \leq i \leq c$, returns also a partial mapping $\lambda_i$ from the \emph{possible initial states} (PIS) of $A_i$ to the \emph{possible last active states} (PLAS) of $A_i$. More precisely, a state pair $(q, \, q')$ is in $\lambda_i$ if it holds $q' \in \delta^A \, (q, \, y_i)$.
\par
To complete parallel recognition, we need to \emph{join} the mappings $\lambda_i$ for all the chunks. Such an operation does not depend on the string length $n$, and for brevity we only describe the serial join.
\par
For the first chunk, the possible initial state set is $\text{PIS}_1 = I_1^A = \set{q_0}$, the mapping is $q_0 \stackrel {\lambda_1} \mapsto \delta^A \, (q_0, \, y_1)$, and consequently the possible last  active states are $\text{PLAS}_1 = \lambda_1 \, (\text{PIS}_1)$.
\par
For every other chunk $y_i$, with $2 \leq i \leq c$, the set $\text{PIS}_i \subseteq I_i^A=Q$  contains all the states $q$ such that $\delta^A \, (q, \, y_i)$ is defined, and is represented by the mapping $\lambda_i$ such that for each $q \in \text{PIS}_i$ it holds $q \stackrel {\lambda_i} \mapsto \delta^A \, \big( q, \, y_i \big)$. We intersect the possible initial states with the possible last states of the upstream CA and we apply the mapping, so as to obtain the possible last states:
\[
\text{PLAS}_i = \lambda_i \, \big( \text{PLAS}_{i-1} \, \cap \, \text{PIS}_i \big) \qquad \text{where $\text{PIS}_i = Q$}
\] 	
The last CA accepts by condition $\text{PLAS}_c \, \cap \,  F \not = \emptyset$, i.e., the possible last active states $\text{PLAS}_c$ include a final state.
\par
\begin{figure}[h]
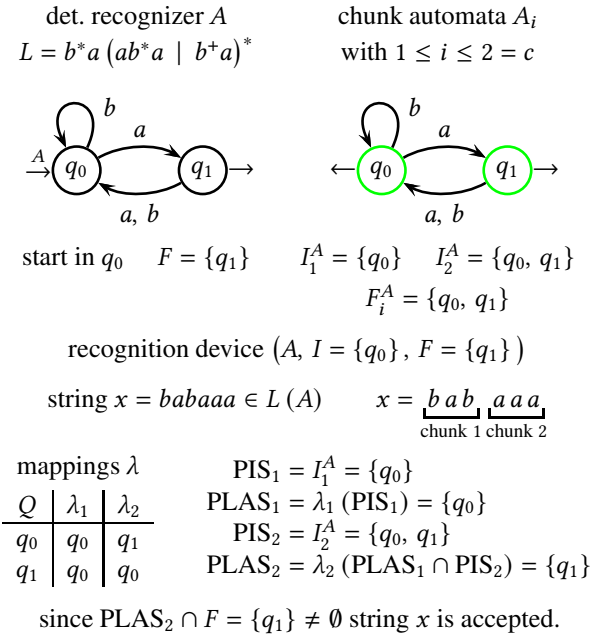

	\begin{center}
		\begin{tabular}{c@{\hspace{0.625cm}}c}
			det. recognizer $A$ & chunk automata $A_i$ \\[0.05cm]
			$L = b^\ast a \, \big(a b^\ast a \, \mid \, b^+ a \big)^\ast$ & with $1 \leq i \leq 2=c$
			\\[1.0cm]
			\begin{automaton}{nodesep=0pt, colsep=1.0cm}
				\circlenode 0 {$q_0$} & \circlenode 1 {$q_1$}
				
				\nput[labelsep=0pt] {170} 0 {$\overset A \rightarrow$}
				
				\nput[labelsep=0pt] {0} 1 {$\rightarrow$}
				
				\nccurve[angleA=65, angleB=115, ncurv=6] 0 0 \nbput[npos=0.25]{$b$}
				
				\ncarc 0 1 \naput{$a$}
				
				\ncarc 1 0 \naput{$a, \, b$}
			\end{automaton}
			&
			\begin{automaton}{nodesep=0pt, colsep=1.0cm}
				\circlenode[linecolor=green] 0 {$q_0$} & \circlenode[linecolor=green]  1 {$q_1$}
				
				\nput[labelsep=0pt] {180} 0 {$\leftarrow$}
				
				\nput[labelsep=0pt] {0} 1 {$\rightarrow$}
				
				\nccurve[angleA=65, angleB=115, ncurv=6] 0 0 \nbput[npos=0.25]{$b$}
				
				\ncarc 0 1 \naput{$a$}
				
				\ncarc 1 0 \naput{$a, \, b$}
			\end{automaton} \\[0.75cm]
			start in $q_0$ \quad $F = \set{q_1}$ & $I_1^A = \set{q_0}$ \quad $I_2^A =\set{q_0, \, q_1}$ \\[0.125cm]
			& $F_i^A = \set{q_0, \, q_1}$ \\[0.25cm]
			\multicolumn{2}{c}{recognition device $\big(A, \, I = \set{q_0}, \, F = \set{q_1}\big)$}
		\end{tabular}
		\par
		\vspace{0.25cm}
		string $x = babaaa \in L \, (A) \qquad x = \underbracket[0.75pt]{\, b \, a \, b \, }_\text{chunk $1$} \underbracket[0.75pt]{ \, a \, a \, a \, }_\text{chunk $2$}$
		\par
		\vspace{0.125cm}
		$\begin{array}{c|c|c}
			\multicolumn{3}{c}{\text{mappings $\lambda$}} \\[0.1cm]
			Q & \lambda_1 & \lambda_2 \\ \hline
			q_0 & q_0 & q_1 \\
			q_1 & q_0 & q_0
		\end{array}
		\qquad
		\arraycolsep=0.0cm\begin{array}{r@{\hspace{0.1cm}}c@{\hspace{0.1cm}}l}
			\text{PIS}_1 & = & I_1^A = \set{q_0} \\
			\text{PLAS}_1 & = & \lambda_1 \, (\text{PIS}_1) = \set{q_0} \\
			\text{PIS}_2 & = & I_2^A = \set{q_0, \, q_1} \\
			\text{PLAS}_2 & = & \lambda_2 \, (\text{PLAS}_1 \cap \text{PIS}_2) = \set{q_1}
		\end{array}$
		\par
		\vspace{0.25cm}
		since $\text{PLAS}_2 \cap F  = \set{q_1} \not = \emptyset$  string $x$ is accepted.
	\end{center}
\caption{\small CSDPA device using DFA. In the CA (top right) all states are initial and final. To recognize the two-chunk input $bab \cdot aaa$, nine transitions are done;  CA $A_2$ executes two $3$-step runs scanning the entire chunk. The join phase (bottom) computes $\text{PLAS}_2$.} \label{fig:Holubmethod}
\end{figure}

An example that uses a deterministic CA is shown in Fig.~\ref{fig:Holubmethod}. The values of the sets PIS and PLAS for the two-chunk input string $bab \cdot aaa$ are listed. Notice that in the example the worst possible speculation overhead occurs, though this is not the case in general.
\par
\noindent
In parallel \emph{deterministic} operation, the overall number of transitions is  bounded by  $T_D = \sum_{i=1}^c \big( \vert \, y_i \, \vert \times \vert \, I_i^A \, \vert \big) \leq n \times \vert \, Q \, \vert$, since each  CA is a  multi-entry machine and some runs may prematurely terminate in error \cite{DBLP:conf/asplos/MytkowiczMS14, DBLP:journals/ijpp/KoJHB14, FuZheLi2017}. The factor $\vert \, Q \, \vert$ is the \emph{speculation cost}, caused by the need  to  speculate on the starting states of CAs. In the \emph{nondeterministic} case, the number of transitions $T_N$  may be higher, similarly to the serial operation.
\par
The CSDPA scheme has inspired many implementations for  various parallel computing architectures. Each CA  is assigned a processing unit that serially executes the number $\vert \, Q \, \vert$ of runs imposed by speculation. As said, sometimes NFAs have been used as a more compact replacement  to reduce  speculation in particular benchmarks, but in general the  benefits are  null or even negative.
\par
We present a new general technique to reduce the speculation cost by combining the  size reduction of a nondeterministic machine with the speed of a deterministic one.
\section{Reduced-interface device (RID)} \label{sec:reduced-interface-device}
Our \emph{reduced-interface device} (RID)  is based on a series of deterministic CAs as the classic device, but with  one important difference.
The CA initial states -- here called \emph{interface} states -- are a typically  smaller subset of the entire state set, since they  exactly mirror the states of an NFA for language $L$. The possible last active states $\text{PLAS}_i$ reached by the $(i-1)$-th CA are still mapped onto the possible initial states $\text{PIS}_{i + 1}$ of the $i$-th CA by an efficient \emph{interface function}. In parallel operation, after all CAs have finished, the input is accepted by the join phase if all the interface mappings are consistent, similarly to the classic case. The RID is a new version of the CSDPA scheme based on DFA, and essentially all the optimizations proposed in past implementations remain possible.
\subsection{Construction of the chunk automaton} \label{subsec:construction}
\begin{figure}[h]
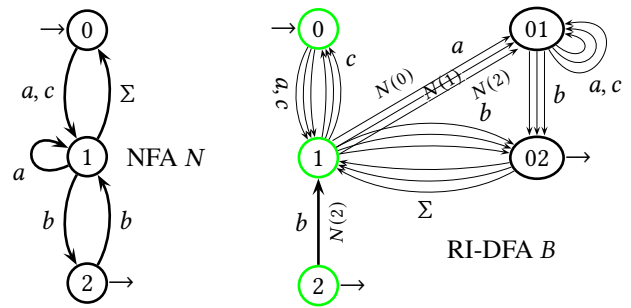

	\begin{center}
		\tabcolsep=0.0cm
		\begin{tabular}{c@{\hspace{2.25cm}}c}
			\begin{automaton}{nodesep=0pt, rowsep=1.125cm, colsep=0.75cm, border=0.0cm}
				\circlenode 0 {$0$} \\
				\circlenode 1 {$1$} \\
				\circlenode 2 {$2$}

				\nput[labelsep=0pt] {180} 0 {$\rightarrow$}
				
				\nput[labelsep=0.25cm] {0} 1 {NFA $N$}
				
				\nput[labelsep=0pt] {0} 2 {$\rightarrow$}
				
				\ncarc[arcangle=-30] 0 1 \nbput{$a, c$}
				
				\nccurve[angleA=-155, angleB=155, ncurv=6] 1 1 \naput[npos=0.375]{$a$}
				
				\ncarc[arcangle=-30] 1 2 \nbput{$b$}
				
				\ncarc[arcangle=-30] 2 1 \nbput{$b$}
				
				\ncarc[arcangle=-30] 1 0 \nbput{$\Sigma$}
			\end{automaton}
			&
			\begin{automaton}{nodesep=0pt, rowsep=1.125cm, colsep=2.25cm, border=0.0cm}

				\circlenode[linecolor=green] 0 {$0$} & \ovalnode {01} {$01$} \\

				\circlenode[linecolor=green] 1 {$1$} & \ovalnode {02} {$02$} \\

				\circlenode[linecolor=green] 2 {$2$}
				
                \nput[labelsep=0pt] {180} 0 {$\rightarrow$}

				\nput[labelsep=1.5cm] {15} 2 {RI-DFA $B$}
				
				\nput[labelsep=0pt] {0} 2 {$\rightarrow$}
				
				\nput[labelsep=0pt] {0} {02} {$\rightarrow$}
				
				\ncarc[arcangle=-30, linewidth=0.2pt] 0 1 \nbput[nrot=:U]{$a, c$}

				\ncarc[arcangle=-20, linewidth=0.2pt] 0 1

                \ncarc[arcangle=-10, linewidth=0.2pt] 0 1
				
				\ncarc[arcangle=-30, linewidth=0.2pt] 1 0 \nbput[npos=0.8]{$c$}

				\ncarc[arcangle=-20, linewidth=0.2pt] 1 0

				\ncarc[arcangle=-10, linewidth=0.2pt] 1 0 				
				
                \ncline[offset=-0.1, nodesepA=0.0, nodesepB=0.0, linewidth=0.2pt] {1} {01} \nbput[npos=0.8, nrot=:U] {\scriptsize $N(2)$}
                \ncline[offset=0, nodesepA=0.0, nodesepB=0.0, linewidth=0.2pt] {1} {01}   \ncput[npos=0.6, nrot=:U] {\scriptsize $\psframebox[framesep=0pt, linewidth=0pt, linecolor=white, fillcolor=white]{N(1)}$}
                \ncline[offset=0.1, nodesepA=0.0, nodesepB=0.0, linewidth=0.2pt] {1} {01} \naput[npos=0.4, nrot=:U] {\scriptsize $N(0)$}  \naput[npos=0.75]{$a$}
				
				\ncarc[arcangle=30, linewidth=0.2pt] 1 {02}  \naput[npos=0.8]{$b$}
				\ncarc[arcangle=20, linewidth=0.2pt] 1 {02}
				\ncarc[arcangle=10, linewidth=0.2pt] 1 {02}

				\ncarc[arcangle=30,linewidth=0.2pt] {02} 1  \naput[npos=0.5]{$\Sigma$}
				\ncarc[arcangle=20,linewidth=0.2pt] {02} 1
				\ncarc[arcangle=10,linewidth=0.2pt] {02} 1
				
				\nccurve[angleA=-60, angleB=10, ncurv=5.5, linewidth=0.2pt] {01} {01} \nbput[npos=0.35]{$a, c$}
				\nccurve[angleA=-50, angleB=-0, ncurv=5, linewidth=0.2pt] {01} {01}
				\nccurve[angleA=-40, angleB=-10, ncurv=5, linewidth=0.2pt] {01} {01}
				
				\ncline[offset=-0.1, nodesepA=0.0, nodesepB=0.0, linewidth=0.2pt] {01} {02}
				\ncline[offset=0, nodesepA=0.0, nodesepB=0.0, linewidth=0.2pt] {->} {01} {02}
				\ncline[offset=0.1, nodesepA=0.0, nodesepB=0.0, linewidth=0.2pt] {->} {01} {02} \naput{$b$}
				
				\ncline 2 1 \naput{$b$} \nbput[npos=0.5, nrot=:U]{\scriptsize $N(2)$}
			\end{automaton}
		\end{tabular}
	\end{center}
\caption{\small Left: the given NFA $N$. Right: the RI-DFA $B$ obtained by incrementally adding to $N(0)$ first $N(1)$ and then $N(2)$. The states $I^B = \big\{ \set{0}, \set{1}, \set{2} \big\}$ coloured in {\green green} act as initial.} \label{fig:RI-DFAconstr-ex}
\end{figure}

We describe the construction of the  CA, denoted $B$ (instead of $A$), and of the chunk interface function,  called \interfun. The RID device  is specified by the CA $B$, by the interface function $\interfun$, and by the initial and final sets $ \set{q_0}$ and $F^{\text{RID}}$. To construct the CA we start from an NFA $N$ for language $L$. In practice such NFAs are available from benchmarks of automata or can be constructed from a benchmark of REs by means of a standard $\text{RE} \to \text{NFA}$ translator. We use the following FAs:
\begin{itemize}[leftmargin=*]
\item The NFA $N = \big( Q_N, \, \Sigma, \, \rho, \, q_0, \, F \big)$ has a state set $Q_N = \set{q_0, \, \ldots, \, q_{\ell - 1} }$, with $\ell \geq 1$ states, initial state $q_0$, transition relation
$\rho$ and final state set $F$.
\item The \emph{reduced-interface deterministic} FA (RI-DFA), denoted $B = \big( P, \, \Sigma, \, \delta^B, \, I^B, \, F^B \big)$, is a multi-entry machine derived from NFA $N$. Device $B$
has a state set $P = \set{ \ldots, \, p_i, \, \ldots }$, transition function $\delta^B$, initial and final state sets $I^B$ and $F^B$.
\end{itemize}
Our RI-DFA is a MDFA since the set $I^B$ contains multiple initial states, i.e., a nondeterministic feature, but  the transition graph is deterministic ($\delta^B$ is a function); see also Sect.~\ref{sec:introduction} -- \emph{theoretical contributions}. The state sets $Q_N$ of $N$ and $P$ of the CA $B_i$ are different, the latter being larger (but it will be reduced in Sect.~\ref{subsec:furtherReduction}). Intuitively the RI-DFA can be obtained by applying the textbook-based powerset construction to the NFA, by enumerating all the subsets of NFA states $Q_N$ and their transitions, determinizing and removing those unreachable from a singleton state. Our construction is more refined, and proceeds incrementally:
\begin{itemize}[leftmargin=*]
\item incrementally apply the $\text{NFA} \to \text{DFA}$ powerset algorithm for $\vert \, Q_N \, \vert = \ell$ times, each time starting from a singleton initial state, as follows:
\begin{itemize}
\item $N(q_0) \asgn$ powerset machine for $N$ with initial state $q_0$
\item $N(q_1) \asgn N(q_0) \; \cup$ additional states and transitions reachable from state $q_1$
\item continue in this way from states $q_2$, $q_3$, etc
\item $N(q_{\ell - 1}) \asgn N(q_{\ell - 2}) \; \cup$ additional states and transitions reachable from state $q_{\ell - 1}$
\end{itemize}
\item the state set $P$ is the union of state sets $N(q_0) \ldots N(q_{\ell - 1})$
\item the initial  states  of  $B$ are $I^B = \big\{ \set{q_0},  \ldots, \set{q_{\ell -1}} \big\} \equiv Q_N$
\item for the RID, the initial state  is $\set{q_0}$ and the final states  $F^{\text{RID}}$ are the union of the final states  of $N(q_0) \ldots N(q_{\ell - 1})$
\end{itemize}
The construction is shown in Fig.~\ref{fig:RI-DFAconstr-ex} for the  example of Fig.~\ref{fig:RIDintroductory}. Our construction $\text{NFA} \to \text{DFA}$ is quite efficient, and some quantitative measurements are reported in Sect.~\ref{sec:experimental-results}.
\par
The state set of RI-DFA includes (as initial) all the states of the given NFA $N$. The state set of RI-DFA and the one of the classic powerset DFA equivalent to $N$, are instead incomparable. In fact, the $\ell$ powerset applications in the construction above may create states absent when the NFA $N$ is determinized in one shot by a single powerset application, and the opposite may also happen. Quite often in practice, the two devices share several states.
\par
In  Fig.~\ref{fig:RI-DFAconstr-ex}, the  powerset machines $N(0)$ and $N(1)$ are identical and have the state set $\big\{ \set{0}, \set{1}, \set{0,1}, \set{0, 2} \big\}$ (state names are shortened, i.e., $q_0$, $q_1$, etc. become $0$, $1$, etc.). Machine $N(2)$ differs from the preceding one only  by the state $\set{2}$ and the transition $\set{2} \xrightarrow b \set{1}$. Thus, the states of the RI-DFA are $P = \big\{\set{0}, \set{1}, \set{2}, \set{0, 1}, \set{0, 2} \big\}$ and comprise the \emph{singleton} states $\set{0}$, $\set{1}$ and $\set{2}$, which act as initial, and the \emph{aggregate} states $\set{0, 1}$ and $\set{0, 2}$. When the relation of an RI-DFA state to the NFA states  represented is not needed, we prefer to denote a state as $p \in P$.
\par\noindent
The number of CA initial states for RI-DFA is the same as for the CSDPA using nondeterministic CAs, and is typically smaller than for deterministic CAs. We anticipate that it can be further reduced (see Sect.~\ref{subsec:furtherReduction}).
\subsection{Parallel operations and interface function} \label{subsec:function}
In parallel operation the behavior of RID essentially conforms  to the CSDPA scheme, and we only focus on the differences.
All CAs $B_i$ start in parallel.  Each  $B_i$ scans its  chunk $y_i$, for each initial state $p \in I_i^B \equiv Q_N$, and of course  $I_1^B = \big\{ \set{q_0} \big\}$. Upon termination, each $B_i$  returns the partial mapping $\lambda_i$ that, from each \emph{possible initial state} (PIS) of $B_i$, gives the \emph{possible last active states} (PLAS): a state pair $(p, \, p')$ is in $\lambda_i$ if it holds $p' = \delta^B \, (p, \, y_i)$.
\par\noindent
\textbf{\emph{Join phase}.} All the consecutive mappings  $\lambda_i$  and $\lambda_{i+1}$ are  joined to check that at least one sequence of chunk runs is consistent.  The RID join operation differs from the CSDPA one by the way it maps the set PLAS of a CA to the set PIS of the downstream CA, by using the \emph{interface function} $\interfun$  next defined. The function argument is  a subset $S \subseteq P$ of states and, for each state $p \in S$ with $p =\set{ \ldots, \, q, \, \ldots }$, where $q \in Q_N$, the result $\interfun \, (p)$ is  the  summation  of  the singleton states $\set{q}$ and is a subset of the initial  set $I^B$.  More precisely $\interfun \colon \wp \, (P) \to \wp \, (I^B)$ with $\interfun \, (\text{PLAS}) = \bigcup_{p \, \in \, \text{PLAS}} \big\{ \set {q} \; \vert \quad q \in p \, \big\}$, where $q \in Q_N$ is an NFA state. The reason for introducing this function \interfun will become clear in Sect.~\ref{subsec:furtherReduction}. Then:
\begin{itemize}[leftmargin=*]
\item For the first chunk, $\text{PIS}_1 = \set{p_0}$ with $p_0 = \set{q_0}$, the mapping is $p_0 \stackrel {\lambda_1} \mapsto \delta^B \, (p_0, \, y_1)$, and for $B_1$ $\text{PLAS}_1 = \lambda_1 \, \big( \text{PIS}_1 \big)$.
\item For every other chunk with $2 \leq i \leq c$, the set $\text{PIS}_i \subseteq I_i^B$ (now $I_i^B$ is a subset of $P$) contains the initial states $p \in I_i^B$ such that $\delta^B \, \big( p, \, y_i \big)$ is defined, and is represented by the mapping $\lambda_i$ such that for each $p \in \text{PIS}_i$ it holds $p \stackrel {\lambda_i} \mapsto \delta^B \, \big( p, \, y_i \big)$.
We obtain the set PLAS of  $B_i$ by intersecting set $\text{PIS}_{i}$ with the upstream set $\text{PLAS}_{i-1}$ through the interface function:
$\text{PLAS}_i = \lambda_i \, \big( \interfun \, \big( \text{PLAS}_{i-1} \big) \, \cap \, \text{PIS}_i \big)	$.
\item	
	 RID recognizes the input if the last CA passes the acceptance condition:
	$
	\text{PLAS}_c \, \cap \,  F^{\text{RID}} \not = \emptyset
	$,
	i.e., at least one of the possible last active states $\text{PLAS}_c$ is final for the RID.
\end{itemize}
The crucial differences with respect to the DFA-based  device  are that (\emph{i}) device RID, through the interface function \interfun, remaps the possible last active states PLAS of the upstream CA $B_{i-1}$ onto the possible initial states PIS of the downstream CA $B_i$, and that (\emph{ii}) set PIS is a subset of the set of the initial states $I^B$, which in turn is a (potentially much smaller) subset of the whole state set $P$ of the CA.
\par
\begin{figure}[tb]
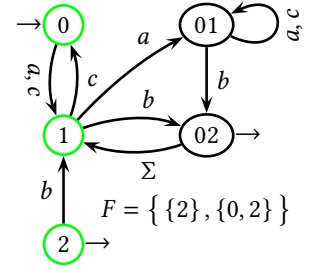

\begin{flushleft}
\tabcolsep=0.0cm
\begin{tabular}{m{4.25cm}@{\hspace{0.5cm}}c}
\normalsize
\textbf{run / mapping of chunk $1$}:
\small
\par
$\set{0} \xrightarrow a \set{1} \xrightarrow a \set{0, 1} \xrightarrow b \set{0, 2}$
\par
$\set{0} \stackrel {\lambda_1} \mapsto \set{0, 2}$
\par
\vspace{0.2cm}
\normalsize
\textbf{runs / mapping of chunk $2$}:
\small
\par
$\set{0} \xrightarrow c \set{1} \xrightarrow a \set{0, 1} \xrightarrow b \set{0, 2}$
\par
$\set{1} \xrightarrow c \set{0} \xrightarrow a \set{1} \xrightarrow b \set{0, 2}$
\par
$\set{0} \stackrel {\lambda_2} \mapsto \set{0, 2}$ \quad $\set{1} \stackrel {\lambda_2} \mapsto \set{0, 2}$
&
\scalebox{1.0}{\begin{automaton}{nodesep=0pt, rowsep=0.875cm, colsep=1.25cm, border=0.0cm}

			\circlenode[linecolor=green] 0 {$0$} & \ovalnode {01} {$01$} \\

			\circlenode[linecolor=green] 1 {$1$} & \ovalnode {02} {$02$} \\

			\circlenode[linecolor=green] 2 {$2$} & \pnode{F}
			
			\nput[labelsep=0pt] {180} 0 {$\rightarrow$}
			
			\nput[labelsep=0pt] {0} 2 {$\rightarrow$}
			
			\nput[labelsep=0.35cm] {95} F {$F = \big\{ \set{2}, \set{0, 2} \big\}$}
			
			\nput[labelsep=0pt] {0} {02} {$\rightarrow$}
			
			\ncarc[arcangle=-20] 0 1 \nbput[nrot=:U]{$a, c$}
			
			\ncarc[arcangle=-20] 1 0 \nbput{$c$}
			
			\ncarc[arcangle=10] 1 {01} \naput[npos=0.75]{$a$}
			
			\ncarc[arcangle=20] 1 {02} \naput[npos=0.65]{$b$}
			
			\ncarc[arcangle=20] {02} 1 \naput[npos=0.35]{$\Sigma$}
			
			\nccurve[angleA=-25, angleB=25, ncurv=6] {01} {01} \nbput[npos=0.5, nrot=:U]{$a, c$}
			
			\ncline {01} {02} \naput{$b$}
			
			\ncline 2 1 \naput{$b$}
\end{automaton}} \\
\multicolumn{2}{m{8cm}}{
\small
\vspace{0.1cm}
\textbf{chunk $1$}: $\text{PLAS}_1 = \big\{ \set{0, 2} \big\}$ and $\interfun \, \big( \big\{ \set{0, 2} \big\} \big) = \big\{ \set{0}, \set{2} \big\}$
\par
\vspace{0.1cm}
\textbf{chunk $2$}: $\text{PIS}_2 = \big\{ \set{0}, \set{1} \big\}$ and $\text{PLAS}_2 = \big\{ \set{0, 2} \big\}$, thus
\par
at last $\text{PLAS}_2 \cap F = \big\{ \set{0, 2} \big\} \cap \big\{ \set{2}, \set{0, 2} \big\} = \big\{ \set{0, 2} \big\} \neq \emptyset$}
\end{tabular}
\end{flushleft}
\caption{\small NFA, runs of CAs $B_1$ and $B_2$, and interface function \interfun.} \label{fig:RI-DFAmethod}
\end{figure}

A case with two chunks is shown in Fig.~\ref{fig:RI-DFAmethod} for the  example of Fig.~\ref{fig:RIDintroductory}.
The input  $aabcab$ is split into $aab$ and $cab$. Processing  $aab$ computes $\text{PLAS}_1 = \big\{ \set{0,2} \big\}$ and $\interfun \, (\text{PLAS}_1) = \big\{ \set{0}, \set{2} \big\}$. The set of possible initial states for  chunk $cab$ is    $\text{PIS}_2 = \big\{ \set{0}, \set{1} \big\}$, therefore $\interfun \, \big( \text{PLAS}_{1} \big) \, \cap \, \text{PIS}_2 = \big\{ \set{0} \big\}$.
Hence  $\text{PLAS}_2 = \lambda_2 \, \big( \interfun \, \big( \text{PLAS}_{1} \big) \, \cap \, \text{PIS}_2 \big) = \big\{ \set{0, 2} \big\} $, which includes the final state $\set{0, 2}$, thus the input is accepted.
\par
\subsection{Correctness of the RID} \label{subsec:correctness}
\begin{theorem}[Correctness] \label{th:correctness}
The \text{\rm RID} accepts the same language as the \text{\rm NFA} $N$.
\end{theorem}
\noindent
We  define some functions to be used in the proof of Th.~\ref{th:correctness}. For the NFA $N = \big( Q_N, \, \Sigma, \, \rho, \, q_0, \, F \big)$, we extend the transition function $\rho \colon Q_N \, \times \, \Sigma \to \wp \, (Q_N) $ to all strings as $\rho^\ast \colon Q_N \, \times \, \Sigma^\ast \to \wp \, (Q_N) $ in the obvious way, i.e., $\rho^\ast \, (q, \, \varepsilon) = \set{q}$ and $\rho^\ast \, (q, \, x \cdot a) = \bigcup_{q' \, \in \, \rho^\ast \, (q, \, x)} \, \rho \, (q', \, a)$, and for simplicity we write $\rho$ in place of $\rho^\ast$. Hence $\rho \, (q, \, x)$ denotes the set of states reached by NFA $N$ from state $q$ after reading string $x$.
For any PLAS of a RID, we introduce the function $\mathit{Nst}$ that returns the union of all the states (of $N$) present in some element $p \in P$  of PLAS, i.e., $\mathit{Nst} \, (\text{PLAS}) = \bigcup_{p \, \in \, \text{PLAS}} \, p$. For any set of states $S$, the function $\mathit{sgl} \, (S)$ returns the set of singletons that contain the elements of $S$, i.e., $\mathit{sgl} \, (S) = \big\{ \set{s} \; \vert \quad s \in S \big\}$.
\begin{lemma} \label{lem:1}
The set of the states of the \text{\rm NFA} $N$ that are included in the elements of each set $\text{\rm PLAS}_i$, is equal to the set of the states reached by $N$ after reading $y_1 \ldots y_i$, i.e., for all $i$ with $1 \leq i \leq c$ it holds $\mathit{Nst} \, (\text{\rm PLAS}_i) = \rho \, (q_0, \, y_1 \ldots y_i)$.
\end{lemma}
\begin{proof}
By induction on the chunk index $i$:
\begin{itemize}[leftmargin=*]
\item For chunk $y_1$, it holds $\text{PLAS}_1 = \delta^B \, (p_0, \, y_1) = \delta^B \, \big( \set{q_0}, \, y_1 \big)$
and, by the definition of $\delta_B$ for RI-DFA, it follows that $\mathit{Nst} \, (\text{PLAS}_1) = \rho \, (q_0, \, y_1)$.
\item For chunk $y_i$ with $2 \leq i \leq c$, inductively assume that
$\mathit{Nst} \, (\text{PLAS}_{i-1})$ $=$ $\rho \, (q_0, \, y_1 \ldots y_{i-1})$. Then it holds:
\begin{eqnarray*}
\text{PIS}_i & = & \big\{ \set{q} \; \vert \quad \text{$\delta^B \, \big( \set{q}, \, y_i \big)$ is defined} \big\} \\
& = & \big\{ \set{q} \; \vert \quad \rho \, (q, \, y_i) \neq \emptyset \, \big\}
\end{eqnarray*}
and, by the inductive assumption:
\begin{eqnarray*}
\interfun \, (\text{PLAS}_{i-1}) & = & \bigcup_{\, p \, \in \, \text{PLAS}_{i-1}} \, \big\{ \set{q} \; \vert \quad q \in p \, \big\} \\
& = & \mathit{sgl} \, \big( \mathit{Nst} \, (\text{PLAS}_{i-1}) \big) \\
& = & \mathit{sgl} \, \big( \rho \, (q_0, \, y_1 \ldots y_{i-1}) \big)
\end{eqnarray*}
Thus it is $\interfun \, (\text{PLAS}_{i-1}) \cap \text{PIS}_i = \mathit{sgl} \, \big( \rho \, (q_0, \, y_1 \ldots y_{i-1}) \big) \, \cap \, \big\{ \set{q} \; \vert \quad \rho \, (q, \, y_i) \neq \emptyset \, \big\}$, $\text{PLAS}_i = \lambda_i \, (\interfun \, (\text{PLAS}_{i-1}) \cap \text{PIS}_i) = \set{\delta^B \, \big( \set{q}, \, y_i \big) \; \vert \quad \set{q} \in \big(\interfun \, (\text{PLAS}_{i-1}) \cap \text{PIS}_i\big)}$, whence:
\begin{align*}
\mathit{Nst} \, (\text{PLAS}_i) & = \set{q \; \vline \quad \begin{array}{l} \exists \, q' \, \big(q' \in \rho \, (q_0, \, y_1 \ldots y_{i-1}) \\ \land \; q \in \rho \, (q', \, y_i)\big) \end{array}} \\
& = \big\{ q \; \vert \quad q \in \rho \, (q_0, \, y_1 \ldots y_i) \big\} \\
& = \rho \, (q_0, \, y_1 \ldots y_i)
\end{align*}
\end{itemize}
\end{proof}
\begin{figure}[b]
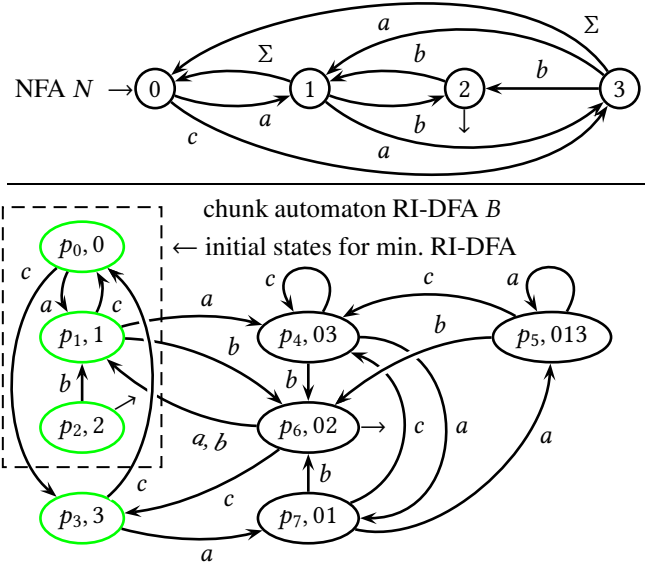

\begin{center}
\vspace{0.625cm}
\hspace{1.375cm}
	\begin{automaton}{nodesep=0pt, rowsep=0.0cm, colsep=1.5cm, border=0.0cm}
		\circlenode 0 {$0$} &
		\circlenode 1 {$1$} &
		\circlenode 2 {$2$} &
		\circlenode 3 {$3$}

		\nput[labelsep=0pt] {180} 0 {NFA $N$ \ $\rightarrow$}
		
		\nput[labelsep=0pt] {-90} 2 {$\downarrow$}
		
		\ncarc[arcangle=-20] 0 1 \nbput[npos=0.75]{$a$}
		
		\nccurve[angleA=-40, angleB=-125, ncurvA=0.5, ncurvB=0.625] 0 3 \nbput[npos=0.1]{$c$}
		
		\ncarc[arcangle=-20] 1 0 \nbput[npos=0.25]{$\Sigma$}
		
		\ncarc[arcangle=-20] 1 2 \nbput[npos=0.75]{$b$}
		
		\ncarc[arcangle=-40] 1 3 \nbput[npos=0.25]{$a$}
		
		\ncarc[arcangle=-20] 2 1 \nbput[npos=0.25]{$b$}
		
		\nccurve[angleA=125, angleB=40, ncurvA=0.625, ncurvB=0.5] 3 0 \nbput[npos=0.1]{$\Sigma$}
		
		\ncarc[arcangle=-40] 3 1 \nbput[npos=0.75]{$a$}
		
		\ncline 3 2 \nbput{$b$}
	\end{automaton}
	\vspace{0.75cm}
	\hrule
	\vspace{0.2cm}
    \par
	\qquad chunk automaton RI-DFA $B$
	\begin{automaton}{nodesep=0pt, rowsep=0.5cm, colsep=1.75cm, border=0.00cm}
	\ovalnode[linecolor=green]  0 {$p_0, 0$} \\

	\ovalnode[linecolor=green]  1 {$p_1,1$} &  \ovalnode {03} {$p_4, 03$}  &  \ovalnode {013} {$p_5,013$} \\

	\ovalnode[linecolor=green]  2 {$p_2,2$} &  \ovalnode {02} {$p_6,02$} \\

	\ovalnode[linecolor=green]  3 {$p_3,3$} &  \ovalnode {01} {$p_7, 01$}
	
	\nput[labelsep=-1pt] {30} 2 {\rotatebox{30}{$\rightarrow$}}
	
	\nput[labelsep=0pt] 0 {02} {$\rightarrow$}
	
	\ncarc[arcangle=-30] 0 1 \nbput[npos=0.75]{$a$}
	
	\ncarc[arcangle=-50] 0 3 \nbput[npos=0.1]{$c$}
	
	\ncarc[arcangle=-30] 1 0 \nbput[npos=0.25]{$c$}
	
	\ncarc[arcangle=20] 1 {02} \naput[npos=0.60]{$b$}
	
	\ncarc[arcangle=15] 1 {03} \naput[npos=0.6]{$a$}
	
	\ncarc[arcangle=0] 2 1 \naput{$b$}
	
	\ncarc[arcangle=-15] 3 {01} \nbput[npos=0.6]{$a$}
	
	\ncline {01} {02} \nbput{$b$}
	
	\ncarc[arcangle=-70, ncurv=1.0] {01} {03} \nbput[npos=0.50]{$c$}
	
	\ncarc[arcangle=-50] {01} {013} \nbput[npos=0.75]{$a$}
	
	\ncarc[arcangle=20] {02} 1 \naput[npos=0.25, nrot=:D]{$a,b$}
	
	\ncarc[arcangle=15] {02} 3 \naput[npos=0.4]{$c$}
	
	\nccurve[angleA=60, angleB=120, ncurv=4.5] {03} {03} \nbput[npos=0.80]{$c$}	
	
	\ncarc[arcangle=90, ncurv=1.25] {03} {01} \naput[npos=0.50]{$a$}	
	
	\ncline {03} {02} \nbput[npos=0.40]{$b$}
	
	\nccurve[angleA=60, angleB=120, ncurv=4.5] {013} {013} \nbput[npos=0.8]{$a$}		
		
    \ncarc[arcangle=-30] {013} {03} \nbput[npos=0.50]{$c$}

	\ncarc[arcangle=-20, border=0.05cm] {013} {02} \nbput[npos=0.275]{$b$}	

	\ncarc[arcangle=-50, border=0.05cm] 3 0 \nbput[npos=0.125]{$c$}

\ncbox[linestyle=dashed, linewidth=0.75pt, boxsize=30pt, nodesep=5pt, arcangle=90, border=0pt] {2} {0}

\nbput[npos=0.85] {\rotatebox {0} {$\leftarrow$ initial states for min. RI-DFA}}				
\end{automaton}
\end{center}
\caption{\small Interface minimization  of an RI-DFA. Top: NFA $N$. Bottom: RI-DFA with initial states $p_0$, $p_1$, $p_2$ and $p_3$ (in green). States $p_1$ and $p_3$ are undistinguishable and  state $p_3$ (arbitrarily chosen) is downgraded from initial to non-initial, thus reducing the initial state set to $\set{p_0, \, p_1, \, p_2}$ (dashed box). The content of state $p_1$ has to be updated to $13$ (not shown here), to adjust the interface function $\interfun$.} \label{fig:minRI-DFAmethod}
\end{figure}
\begin{lemma} \label{lem:2} The acceptance condition of \text{\rm RID} holds if and only if the acceptance condition of \text{\rm NFA} $N$ holds, that is:
$\text{\rm PLAS}_c \, \cap \,  F^{\text{\rm RID}} \not = \emptyset \ \iff \
\rho \, (q_0, \, y_1 \ldots y_c) \, \cap \, F \not = \emptyset$.
\end{lemma}
\begin{proof}
We prove the two implications, by using Lemma~\ref{lem:1}:
\begin{itemize}[leftmargin=*]
\item $ \text{PLAS}_c \, \cap \,  F^{\text{RID}} \not = \emptyset \Rightarrow \rho \, (q_0, \, y_1 \ldots y_c) \cap F \not = \emptyset$.
In fact, if $\text{PLAS}_c \, \cap \,  F^{\text{RID}} \not = \emptyset$, some element of $\text{PLAS}_c$ includes a final state of NFA $N$, and so does $\mathit{Nst}\, (\text{PLAS}_c)$, thus $\mathit{Nst} \, (\text{PLAS}_c) \cap F \not = \emptyset$. From Le.~\ref{lem:1}, $\mathit{Nst} \, (\text{PLAS}_c) = \rho \, (q_0, \, y_1 \ldots y_c )$, hence $\rho \, (q_0, \, y_1 \ldots y_c ) \, \cap \, F \not = \emptyset$.
\item $\rho \, (q_0, \, y_1 \ldots y_c) \, \cap \, F \not = \emptyset \Rightarrow \text{PLAS}_c \, \cap \,  F^{\text{RID}} \not = \emptyset$.
Again from Le.~\ref{lem:1}, $\rho \, (q_0, \, y_1 \ldots y_c) = \mathit{Nst} \, (\text{PLAS}_c)$, hence if
$\rho \, (q_0, \, y_1 \ldots y_c)$ includes a final state of NFA $N$, some element of $\text{PLAS}_c$ also includes a final state of NFA $N$, i.e., it is an element of $F^{\text{RID}}$, thus $\text{PLAS}_c \, \cap \,  F^{\text{RID}} \not = \emptyset$.
\end{itemize}
\end{proof}
\noindent
Th.~\ref{th:correctness} follows immediately from the equivalence of the acceptance conditions of RID and NFA $N$ stated in Lemma~\ref{lem:2}.
\subsection{Further reduction of the interface} \label{subsec:furtherReduction}
The program used in our  experimentation (see Sect.~\ref{sec:experimental-results}) implements the RID by using an RI-DFA as chunk automaton, with an additional optimization of the machine constructed in Sect.~\ref{subsec:construction} to further reduce the interface size.  This is achieved by downgrading from initial to non-initial some states the role of which as initial can be taken by another initial state. We explain our minimization, and at the end  we discuss how the resulting automaton  is related with the minimal NFAs and with other types of NFAs of small size.
\par
A  natural but naive idea is to apply to RI-DFA the standard so-called state-partition algorithm for converting a  DFA to the equivalent DFA with minimal number of states. It is well known that each state corresponds to  a class of equivalence for the so-called Nerode relation, a.k.a. language-equivalence relation:  two states are language-equivalent if they recognize the same language. To compute the equivalence classes, the well-known algorithm, omitted for brevity, partitions the states into the maximal classes of undistinguishable states. Our use  of the state-partition algorithm significantly differs from the usual applications. In fact the partition algorithm is intended  for deterministic machines, while RI-DFA has a nondeterministic choice for the first transition. This notwithstanding, the  language-equivalence relation can be  extended  to an RI-DFA machine, since from each state, including the initial ones, the outgoing transitions are deterministic.
\par
We show on the example in Fig.~\ref{fig:minRI-DFAmethod} how to  use the equivalence classes to reduce the number of initial states.
Consider the RI-DFA $B$ in  Fig.~\ref{fig:minRI-DFAmethod}, bottom. Each state $p_i$ is decorated with the set of states of  NFA $N$, e.g., state $p_4$ has the states $0$ and $3$ of $N$. The four states $p_0$, $p_1$, $p_2$ and $p_3$, which correspond to the initial states of $N$, are singletons. Two non-trivial language-equivalent classes  are computed by the state-partition algorithm: $\alpha = \set{p_1, \, p_3}$ and  $\beta = \set{p_4, \, p_5, \, p_7}$. The former is a subset of the initial state set, while the latter is not, and for our purposes we only need to consider class $\alpha$, which we call an \emph{initial-state} equivalence class. Now, the conventional state-minimization algorithm would coalesce the classes $\alpha$ and $\beta$ into two respective states, yet we do not need to do so, as we do not care to reduce the total number of states, but just the initial ones. It is simpler to arbitrarily pick one of the states of the  initial-state  equivalence class $\alpha = \set{p_1, \, p_3}$, and to downgrade the remaining state(s) as non-initial. In Fig.~\ref{fig:minRI-DFAmethod}, we choose $p_1$ and downgrade $p_3$ as non-initial, thus reducing the initial states (enclosed in the dashed box) from $4$ to $3$. We say that state $p_3$ delegates to state $p_1$ its role as initial. The state transition graph is unchanged.
At the same time, the  content of state $p_1$ has to be updated from $1$ to $13$ (not shown in the figure) to record that, if the upstream CA admits state $3$ as last possible,  then the current CA should start (also) in state $3$.
\par
\begin{figure}[tb]
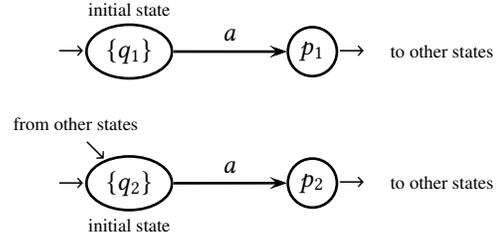
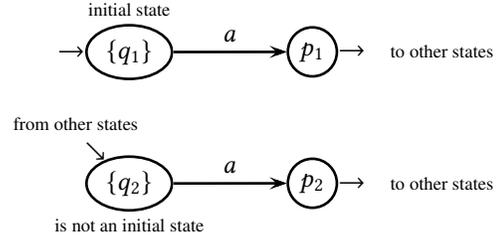
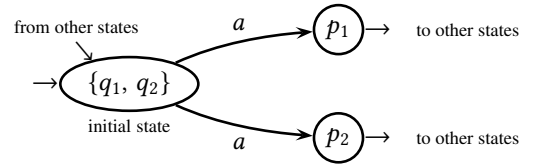

\begin{subfigure}[c]{0.475\textwidth}
\begin{center}
\vspace{0.5cm}
\begin{automaton}{nodesep=0.0cm, rowsep=1.0cm, nodealign=true}
\ovalnode {q1} {$\set{q_1}$} & \circlenode {p1} {$p_1$} & \\
\ovalnode {q2} {$\set{q_2}$} & \circlenode {p2} {$p_2$} &

\nput[labelsep=0.0cm] {180} {q1} {$\to$}

\nput[labelsep=0.1cm] {90} {q1} {\scriptsize initial state}

\nput[labelsep=0.0cm] {180} {q2} {$\to$}

\nput[labelsep=0.1cm] {-90} {q2} {\scriptsize initial state}

\nput[labelsep=0.0cm] {0} {p1} {$\to$ \scriptsize \quad to other states}

\nput[labelsep=0.0cm] {0} {p2} {$\to$ \scriptsize \quad to other states}

\nput[labelsep=0.0cm, rot=135] {135} {q2} {$\leftarrow$}
\nput[labelsep=0.375cm] {115} {q2} {\scriptsize from other states}

\ncline {q1} {p1} \naput{$a$}

\ncline {q2} {p2} \naput{$a$}
\end{automaton}
\end{center}
\vspace{0.375cm}
\caption{Fragment of a non-minimal RI-DFA $B$ (the machine constructed in Sect.~\ref{subsec:construction}), where we assume states $\{q_1\}$ and $\{q_2\}$ to be initial and Nerode-equivalent and consequently also $p_1 \equiv p_2$.} \label{subfig:minimization-part1}
\end{subfigure}
\begin{subfigure}[c]{0.475\textwidth}
\vspace{0.5cm}
\begin{center}
\begin{automaton}{nodesep=0.0cm, rowsep=1.0cm, nodealign=true}
\ovalnode {q1} {$\set{q_1}$} & \circlenode {p1} {$p_1$} & \\
\ovalnode {q2} {$\set{q_2}$} & \circlenode {p2} {$p_2$} &

\nput[labelsep=0.0cm] {180} {q1} {$\to$}

\nput[labelsep=0.1cm] {90} {q1} {\scriptsize initial state}

\nput[labelsep=0.1cm] {-90} {q2} {\scriptsize is not an initial state}

\nput[labelsep=0.0cm] {0} {p1} {$\to$ \scriptsize \quad to other states}

\nput[labelsep=0.0cm] {0} {p2} {$\to$ \scriptsize \quad to other states}

\nput[labelsep=0.0cm, rot=135] {135} {q2} {$\leftarrow$}
\nput[labelsep=0.375cm] {115} {q2} {\scriptsize from other states}

\ncline {q1} {p1} \naput{$a$}

\ncline {q2} {p2} \naput{$a$}
\end{automaton}
\end{center}
\vspace{0.375cm}
\caption{Our minimal RI-DFA $B_{min}$. Notice that  states $q_1$ and $q_2$ are left separate, as well as states $p_1$ and $p_2$, but state $q_1$ is delegated the initial role.} \label{subfig:minimization-part2}
\end{subfigure}
\begin{subfigure}[c]{0.475\textwidth}
\vspace{0.375cm}
\begin{center}
\begin{automaton}{nodesep=0.0cm, rowsep=0.0cm, border=0.0cm, nodealign=true}
& \circlenode {p1} {$p_1$} & \\
\ovalnode {q1q2} {$\set{q_1, \, q_2}$} \\
& \circlenode {p2} {$p_2$} &

\nput[labelsep=0.0cm] {180} {q1q2} {$\to$}

\nput[labelsep=0.1cm] {-90} {q1q2} {\scriptsize initial state}

\nput[labelsep=0.0cm] {0} {p1} {$\to$ \scriptsize \quad to other states}

\nput[labelsep=0.0cm] {0} {p2} {$\to$ \scriptsize \quad to other states}

\nput[labelsep=0.0cm, rot=130] {145} {q1q2} {$\leftarrow$}
\nput[labelsep=0.325cm] {115} {q1q2} {\scriptsize from other states}

\ncarc[arcangle=10] {q1q2} {p1} \naput{$a$}

\ncarc[arcangle=-10] {q1q2} {p2} \nbput{$a$}
\end{automaton}
\end{center}
\vspace{0.25cm}
\caption{If states $q_1$ and $q_2$ were merged, but states $p_1$ and $p_2$ were left separate, then the RI-DFA would become nondeterministic. For the RI-DFA to remain deterministic, also states $p_1$ and $p_2$ should be merged, which for efficiency we prefer not  to do.} \label{subfig:minimization-part3}
\end{subfigure}
\caption{The reduction of  the number of initial states of the RI-DFA is achieved by downgrading a state to non-initial rather than merging it with  the equivalent states. } \label{fig:minimization}
\end{figure}

To explain more visually why  downgrading with delegation is preferable to merging equivalent initial states, in Fig.~\ref{fig:minimization} we sketch the simplest possible case. Fig.~\ref{subfig:minimization-part1} shows a fragment of a RI-DFA machine $B$ (constructed as  in Sect.~\ref{subsec:construction}). Fig.~\ref{subfig:minimization-part2} shows the machine $B_\emph{min}$ that delegates the initial role of state $\set{q_2}$ to state $\set{q_1}$, but does not pay the cost of merging $\set{q_1}$ with $\set{q_2}$ and consequently $p_1$ with $p_2$. On the contrary, the machine in Fig.~\ref{subfig:minimization-part3} loses determinism after merging the two initial states and cannot be used as an efficient CA, unless also the remaining equivalent states are merged. This  for  sure adds a cost to the CA construction, without  guaranteeing    any performance gain for the recognizer.
\par
As said, the minimization algorithm transforms  an RI-DFA $B$ into a machine of the same type, denoted $B_\emph{min}$, identical to $B$ except that one or more initial states of $B$ are no longer such. In the example, state $p_3$ has turned to non-initial, while state $p_1$ has taken over the role of $p_3$ as initial, i.e., its content has become $13$ instead of $1$. In order to use $B_\emph{min}$ as CA, the RID interface function (defined in Sect.~\ref{subsec:function}  ) is adjusted:
\begin{equation*}
\interfun_{\emph{min}} \, (\text{PLAS}) = \Big\{ \, p \in I^{B_{\emph{min}}} \; \vline \quad {\footnotesize \def\arraystretch{1.0} \begin{array}{l} \exists \, \set{q} \in \interfun \, (\text{PLAS}) \ \text{such that} \\
\text{$\set{q} = p$ or $\set{q}$ delegates to $p$} \end{array}} \Big\}
\end{equation*}
We sketch the proof that the device, denoted $\text{RID}_\emph{min}$, is correct, by using the minimized CAs $B_{\emph{min},1} \ldots B_{\emph{min},c}$ and the adjusted function $\interfun$ for the join phase, see the equation above.
\par
The runs speculatively executed by $B_\emph{min}$ do not include those that start from the states of $I^B \setminus I^{B_\emph{min}}$, i.e., those from states that were downgraded from initial to non-initial. However, the interface function $\interfun_\emph{min}$ maps all such downgraded states, possibly included in the set PLAS, to the initial states of  $I^{B_\emph{min}}$ that are their delegates, hence are language-equivalent by construction. Therefore, no RID run useful for acceptance is lost in the $\text{RID}_\emph{min}$ and no new accepting run is introduced, hence the devices RID and $\text{RID}_\emph{min}$ are equivalent.
\par
For any given chunk $y_i$, we figure out how many transitions are saved by the CA $B_{\emph{min}, \, i}$.  Let $I^B \setminus I^{B_\emph{min}}$ be the set of states downgraded from initial to non-initial. Then any  run of $B_i$ that starts in a downgraded state is entirely avoided by $B_{\emph{min}, \, i}$. For the RID of Fig.~\ref{fig:minRI-DFAmethod}, and the chunks $y_1 = caa$ and  $y_2 = aab$, we have the five runs below:
\begin{center} \footnotesize
\vspace{0.0cm}
$\begin{array}{lll}
p_0 \xrightarrow c p_3 \xrightarrow a p_7 \xrightarrow a p_5 &
p_0 \xrightarrow a p_1 \xrightarrow a p_4 \xrightarrow b p_6 &
p_1 \xrightarrow a p_4 \xrightarrow a p_7 \xrightarrow b p_6 \\
p_2 \to \; \dashv \ \text{(exits)} &
p3 \xrightarrow a p_7 \xrightarrow a p_4 \xrightarrow b p_6
\end{array}$
\vspace{0.0cm}
\end{center}
The last run of chunk $2$ does not occur with $\text{RID}_\emph{min}$. Notice that $\text{PLAS}_1 = \set{p_5} = \big\{ \set{ 0, 1, 3} \big\}$, state $\set{3}$ is downgraded to non-initial and its delegate is state $\set{1}$, thus it holds $\interfun_\emph{min} \, (\text{PLAS}_1) = \big\{ \set{0}, \set{1} \big\}$.
\par
A natural question is how the CA constructed by using the initial state reduction procedure in Fig.\ref{fig:minRI-DFAmethod} compares with the CA that would be obtained by the same procedure starting from a state-minimal NFA. Th.~\ref{th:minimality} is immediate to prove by contradiction. It implies that applying initial-state reduction to the RI-DFA of a minimal NFA is useless.
\begin{theorem}[Minimality] \label{th:minimality}
Let $N$ be an \text{\rm NFA} and $N_{min}$ be one of the state-minimal equivalent machines. Let $B$ and $C$ be the \text{\rm RI-DFA}s constructed  by the procedure in Sect.~{\rm\ref{subsec:construction}} starting from $N$ and $N_{\mathit{min}}$, respectively. Let $B_{\mathit{min}}$ be the \text{\rm RI-DFA} constructed from  $B$ by the procedure in Fig.~{\rm\ref{fig:minRI-DFAmethod}}.
For any \text{\rm NFA} $N$, the number of initial states in $C$ is less than or equal to the number of initial states in $B_{\mathit{min}}$.
\end{theorem}
\section{Experimentation results} \label{sec:experimental-results}
To evaluate the parallel recognition algorithm (Sect.~\ref{sec:reduced-interface-device}), we  implemented it as a Java software tool that is available on the public site \ZenodoHREF, and we ran an experimentation campaign. First we outline the software architecture,  then we report  measurements  to show the reduction of the number of CA initial states achieved by RI-DFA vs the DFA and NFA variants of CSDPA for a large FA collection, the reduction of the number of transitions executed by RI-DFA CAs vs the same variants, and eventually the speedup of RID vs the same two variants on a multi-core computer. At last we report the construction times of RI-DFA, which show that the extra cost over the DFA construction is moderate and does not jeopardize practical application.
\par
The tool is coded in Java (rel.~$19$) and  includes: a \emph{generator} of the RI-DFA automaton from  either  an RE or an FA, a \emph{parallel recognizer}  for recognizing  user supplied texts, and a \emph{test driver} to  measure performance. The \emph{generator} supports: $\text{RE} \to \text{NFA}$ and $\text{NFA} \to \text{DFA}$ conversions, DFA minimization and RI-DFA construction, including interface minimization. The \emph{parallel recognizer} supports the CSDPA variants that use NFA, DFA or RI-DFA as CA. Parallel execution is based on the Java Thread model, which builds on Linux  thread. Each thread of the algorithm, i.e., each  $\text{CA}_i$,  is  a Java thread. The recognizer reads the text and sets all the chunk boundaries. Then it runs a thread for each chunk, which  prepares the information for the join phase. Upon  completion of all CAs, the algorithm  joins their results (see Sect.~\ref{subsec:function}).
\par
To run the threads in parallel, the tool creates a thread pool and starts it by means of an \emph{Executor-Service} method, which waits for the termination of all the threads in a pool and collects their results. This method serializes the reach and join phases, which is the only
synchronization requirement. Thus each thread can perform its job at full speed without the penalty of any waiting time. All the  accesses to  CAs and to  chunks are read-only. For brevity we do not  discuss  low-level optimizations for balancing the thread work, since those are neutral w.r.t. the relative speeds of the threads, and since here we do not report absolute execution times. The join phase is serial, not being worthy of parallelization.
\subsection{Benchmarks} \label{subsec:benchmarks}
Choosing data-sets for evaluating FA tools is notoriously difficult and subjected to the risk of introducing application bias caused by the very disparate characteristics of the languages typical of each application area, see e.g.,~\cite{DBLP:conf/iiswc/WaddenDBTGSWBRS16}, where the problem is discussed and a benchmark suite is proposed for a specific range of applications / architectures. Our choice is obviously oriented towards the  assessment of the advantages, if any, of the RI-DFA variant over the other two classic ones.  This can be articulated into  the following questions: how significant is the reduction over CSDPA  for
(\emph{i}) the number of CA initial states, (\emph{ii}) the number of CA state transitions, and (\emph{iii}) the execution time on multi-core architectures.
Clearly, for (\emph{i}) it suffices a collection of FAs, while for  (\emph{ii}) and (\emph{iii})  we also need a set of input texts for each FA.
The  measurements for (\emph{i}), (\emph{ii}) and (\emph{iii}) are respectively presented in Sect.s~\ref{subsec:experimental-minimality}, \ref{subsec:experimental-speculation}  and \ref{subsec:experimental-parallel-recognition}, and are summarized at the end. The benchmarks used and their relevant features are listed in Tab.~\ref{tab:benchmarks}.
\par
\begin{table}[tb]
 	\caption{\small Benchmarks -- public benchmarks are starred.} \label{tab:benchmarks}
 	\begin{center}
 		\tabcolsep=0.125cm\def\arraystretch{0.75}
 		\begin{tabular}{@{\hspace{0.0cm}}p{1.5cm}ccr@{\hspace{0.0cm}}}
 			\emph{name} & \emph{n. of} NFAs & \emph{n. of states} & \emph{max text length} \\ \toprule
 			\textbf{\OndrikHREF}$^\ast$  & $1084$ & $2490$ (avg) & none \\
 			\multicolumn{4}{l}{\text{\footnotesize large collection of NFAs that have different purposes}} \\	\multicolumn{4}{l}{\tiny \Ondrik}
 			\\ \midrule
 			\textbf{bigdata}  & $1$ & $5$ & $13$ Mbyte \\
 			\multicolumn{4}{l}{\text{\footnotesize simple synthetic NFA from a short regular expression}} \\ \midrule
 			\textbf{regexp}  & a series & $k + 1 \geq 1$ & $6$ Mbyte \\
            \multicolumn{4}{p{8.0cm}}{\footnotesize series of synthetic NFAs, the DFAs of which grow exponentially, from the well-known REs $(a \mid b)^\ast a \, (a \mid b)^k$ with parameter $k \geq 0$} \\ \midrule
 			\textbf{\bibleHREF}$^\ast$
 			\footnotesize
 			& $1$ & $16$ & $4$ Mbyte \\
 			\multicolumn{4}{l}{\text{\footnotesize HTML manuscript of The Holy Bible}} \\ \multicolumn{4}{l}{\tiny \bible}
 			\\ \midrule
 			\textbf{\fastaHREF}$^\ast$  & $1$ & $29$ & $765$ Kbyte \\
 			\multicolumn{4}{l}{\text{\footnotesize biometric data consisting of various DNA sequences}} \\
 			\multicolumn{4}{l}{\scalebox{0.9}{\tiny \fasta}}
 			\\ \midrule
 			\textbf{\trafficHREF}$^\ast$ & $1$ & $101$ & $11$ Mbyte
 			\\	\multicolumn{4}{l}{\text{\footnotesize system log file of network traffic records}} \\ \multicolumn{4}{l}{\tiny \traffic}
 			\\ \bottomrule
 		\end{tabular}
 	\end{center}
\end{table}
\noindent
\emph{\OndrikHREF} is a collection of large FAs originating from various applications, such as system modeling and formal verification, without texts. The other data-sets contain one NFA, parametric for \emph{regexp}, and six texts of increasing length up to the value in the rightmost column (of Tab.~\ref{tab:benchmarks}). Data-sets \emph{bigdata} and \emph{regexp} are synthetic. Benchmark \emph{bigdata} is obtained from a collection of random REs and texts, generated by our previous tool \REgenHREF~\cite{DBLP:conf/wia/BorsottiBCM19} and then converted into a NFA by method \cite{DBLP:journals/tc/McNaughtonY60}. The NFA of \emph{regexp} causes an exponential explosion in the equivalent minimal DFA. For \emph{\bibleHREF}, we described the titles of the HTML h$3$ subsections by an RE, thus modeling the file as a long text where some instances of the RE occur. For \emph{\fastaHREF}, we wrote an RE to search a few short DNA subsequences. In \emph{\trafficHREF} a large NFA, obtained from an RE describing \emph{syslog} traces, is applied to a system log file of network traffic records.
\subsection{Results on interface minimality} \label{subsec:experimental-minimality}
The  RI-DFA in  Fig.~\ref{fig:RIDintroductory} has fewer initial states than the minimal DFA, and we address two important questions: how frequently this happens and how large the  state reduction is. In Tab.~\ref{tab:Ondrik} the sizes of the NFA and minimal DFA are compared against the number of initial states of the  RI-DFA for all the FAs  in \emph{\OndrikHREF}, the sizes of which (on the average $2490$ states) are much larger  than for the other benchmarks. Column $3$ (of Tab.~\ref{tab:Ondrik}) shows that all RI-DFAs have fewer initial states than their equivalent DFAs, thus confirming that the frequency of cases  that may  benefit from the RI-DFA approach is  very high. The top value $636$ in the first row says that for more than half of the collection, the reduction of initial states is in the range $0.5$-$0.6$. Remember that the classic DFA-based  variant of CSDPA uses all the DFA states as initial, and that starting  chunk recognition from fewer states reduces speculation. Column $2$ shows that $96.4\%$ of the FAs have smaller size than the minimal DFAs, but does not say whether such nondeterministic FAs are minimal. In the Conclusion (Sect.~\ref{sec:conclusion}), we discuss how to materialize the potential gain from NFA.
\par
\begin{table}[tb]
\caption{\small Distribution of \emph{\OndrikHREF} machines with respect to the number of initial states. The leftmost column lists the intervals of width $0.1$ from $0.5$ to $1.4$. Column NFA: number of machines such that the ratio of the number of NFA states over the number of DFA states falls in the interval.  Column RI-DFA: number of machines such that the ratio of the number of RI-DFA initial states over the number of DFA states falls in the interval.} \label{tab:Ondrik}
 	\begin{center}
 		\tabcolsep=0.0cm\def\arraystretch{0.75}
 		\begin{tabular}{l@{\hspace{1.25cm}}c@{\hspace{1.25cm}}c}
 			\emph{interval $<1$} & NFA & RI-DFA \\ \toprule
 			$0.5$ -- $0.6$   &   $110$  &   $636$ \\ \midrule
 			$0.6$ -- $0.7$   &   $677$  &   $355$ \\ \midrule
 			$0.7$ -- $0.8$   &   $173$  &   $34$  \\ \midrule
 			$0.8$ -- $0.9$   &   $60$   &   $40$  \\ \midrule
 			$0.9$ -- $1.0$   &   $25$   &   $19$  \\ \midrule
 			\emph{subtotal}  &   $1045$ \ ($96.4\%$) &  $1084$ \ ($100\%$) \\ \bottomrule \\
 			\emph{interval $>1$} & NFA & RI-DFA \\ \toprule
 			$1.0$ -- $1.1$   &   $19$  &   none   \\ \midrule
 			$1.1$ -- $1.2$   &   $16$  &   none   \\ \midrule
 			$1.2$ -- $1.3$   &   $3$   &   none   \\ \midrule
 			$1.3$ -- $1.4$   &   $1$   &   none   \\ \midrule
 			\emph{subtotal} & $39$ \ ($3.6\%$) & none \\ \bottomrule
 		\end{tabular}
 	\end{center}
\end{table}
\subsection{Results on speculation overhead} \label{subsec:experimental-speculation}
\begin{figure}[b]
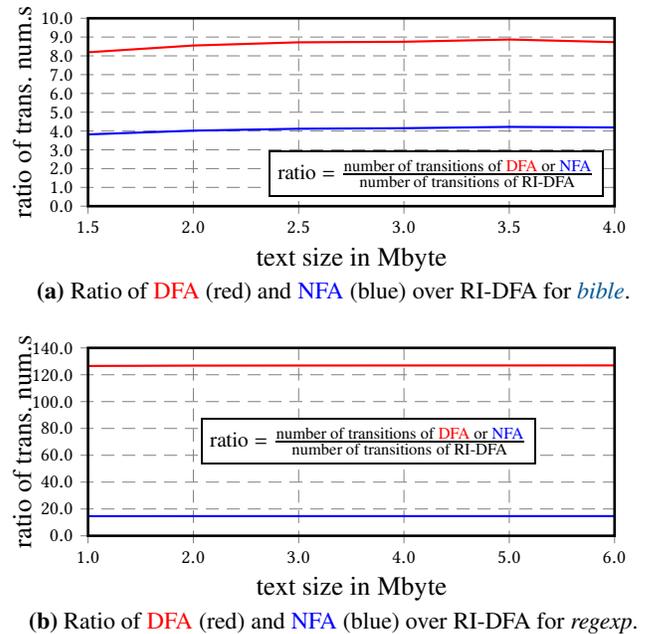

	\begin{subfigure}[c]{0.5\textwidth}
		\begin{center}
			\setpstgraph{gridstyle=dashed, gridwidth=.2pt}
			\rnode{ratio1}{\begin{pstgraph}[xmin=1.5, xmax=4.0, ymin=0, ymax=10, xgriddiv=5, ygriddiv=10](7,2.5)
				\pstfileplot[linecolor=red]{files/bible-trans-det.data.txt}
				\pstfileplot[linecolor=blue]{files/bible-trans-nondet.data.txt}
				\pstlabel{text size in Mbyte}{ratio of trans. num.s}
			\end{pstgraph}}
            \nput[labelsep=-4.6cm] {40} {ratio1} {\fcolorbox{black}{white}{\footnotesize $\text{ratio} = \frac {\text{number of transitions of {\red DFA} or {\blue NFA}}} {\text{number of transitions of RI-DFA}}$}}
		\end{center}
		\caption{Ratio of {\red DFA} (red) and {\blue NFA} (blue) over RI-DFA for \emph{\bibleHREF}.} \label{subfig:bible-transratio}
	\end{subfigure}
	\par
	\vspace{0.5cm}
	\begin{subfigure}[c]{0.5\textwidth}
		\begin{center}
			\setpstgraph{gridstyle=dashed, gridwidth=.2pt}
			\rnode{ratio2}{\begin{pstgraph}[xmin=1, xmax=6, ymin=0, ymax=140, xgriddiv=5, ygriddiv=7](7,2.5)
				\pstfileplot[linecolor=red]{files/regexp-trans-det.data.txt}
				\pstfileplot[linecolor=blue]{files/regexp-trans-nondet.data.txt}
                \pstlabel{text size in Mbyte}{ratio of trans. num.s}
			\end{pstgraph}}
            \nput[labelsep=-5.5cm] {-15} {ratio2} {\fcolorbox{black}{white}{\footnotesize $\text{ratio} = \frac {\text{number of transitions of {\red DFA} or {\blue NFA}}} {\text{number of transitions of RI-DFA}}$}}
		\end{center}
		\caption{Ratio of {\red DFA} (red) and {\blue NFA} (blue) over RI-DFA for \emph{regexp}.} \label{subfig:regexp-transratio}
	\end{subfigure}
	\caption{Transition ratio for benchmarks \emph{\bibleHREF} and \emph{regexp}.} \label{fig:transitionratio}	
\end{figure}

For the five benchmarks in Sect.~\ref{subsec:benchmarks} that have  attached texts, we measured the number of transitions. In fact,  the number of transitions executed by the CAs is  almost directly related to the time speedup for text recognition, to be later reported.
In Fig.~\ref{fig:transitionratio} we show the number of transitions made by the three 	recognizer variants. For brevity, we reproduce only the results for texts  divided in $32$ chunks, which is the mid value between $1$ (serial execution) and $64$, the  number of cores available on our computing platform.
For each benchmark, both the ratio of the numbers of transitions of {\red DFA} over RI-DFA and the ratio of the numbers of transitions of {\blue NFA} over RI-DFA are plotted in {\red red} and {\blue blue}, respectively. We found that these two ratios are close to $1 \pm 10\%$ for the group of benchmarks \emph{bigdata}, \emph{\fastaHREF} and \emph{\trafficHREF} (the pictures are omitted as uninformative), while the ratios are  $\gg 1$ for  group \emph{\bibleHREF} and \emph{regexp}. Let the two benchmark groups be nicknamed \emph{even} and \emph{winning}, respectively. The winning cases are plotted in Fig.s~\ref{subfig:bible-transratio} and \ref{subfig:regexp-transratio}. Notice their almost independence of the text length in the interval plotted. For instance, for \emph{\bibleHREF} the {\red DFA} / RI-DFA transition ratio falls between $8$ and $9$.
\par
We discuss the  difference between the even and winning benchmark groups:   the NFA of, say, \emph{bigdata}  is  far from minimal, being obtained via a basic $\text{RE} \to \text{NFA}$ conversion from a randomly generated RE \cite{DBLP:conf/wia/BorsottiBCM19}. On the other hand, the NFA of \emph{regexp}  is  a classic case of  DFA  state explosion. The low sensitivity of the transition ratios to the text length is presumably explained by the statistics of short-lived  runs, which stop after just a few transitions.
\par
\subsection{Results for parallel recognition} \label{subsec:experimental-parallel-recognition}
For parallel execution, we report the speedup of the RI-DFA variant of CSDPA, i.e., RID, over the other two. We only present comparative time performance evaluations, because absolute times are not at issue here and can be measured by replicating the artifact codes, available on \ZenodoHREF. Our computing platform is a shared-memory multi-core Dell PowerEdge R$7425$ server featuring two AMD EPYC $7551$ $64$-bit CPUs, each with $32$ cores (ISA x$64$), for a total of $2 \times 32 = 64$ identical cores running at a $2.0$ GHz clock frequency. Each core has an L$1$ cache memory of $96$ Kbyte ($64$K for instructions and $32$K for data) and an L$2$ cache memory of $512$ KByte. Each CPU (with $32$ cores) has an L$3$ cache memory of $64$ Mbyte, shared across its $32$ cores. The platform has a main memory of $512$ GByte, shared across all $64$ cores. The  OS is Debian GNU/Linux $6$.$1$.$0$-$13$-amd$64$, with an SMP $6$.$1$.$551$-x$86$-$64$ kernel;  the PREEMPT\_DYNAMIC option is enabled. The same level of compilation optimization was consistently used, and during the measurement campaign the platform was running exclusively our parallel parser, so that the measurements are not affected by external factors.
\par
For each text, the reach phase launches $c = 2, 10, 18, \ldots, 66$ concurrent threads, one per chunk. Since the number of cores is $64$, we can reasonably assume that each thread runs on one core (with the minor exception of case $c = 66$) until termination. To exclude I/O time, the texts are initially loaded into memory and then are processed. For the same benchmarks as in Sect.~\ref{subsec:experimental-minimality}, we measured the recognition time
as a function of two independent variables: text length and number of threads. We select here a  representative sample of our measures.
In all cases the reach phase takes the longer time, while the join phase accounts for less than $1\%$ of the total time.
\par
\begin{figure}[tb]
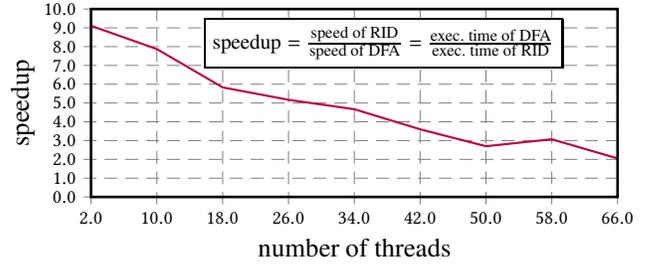
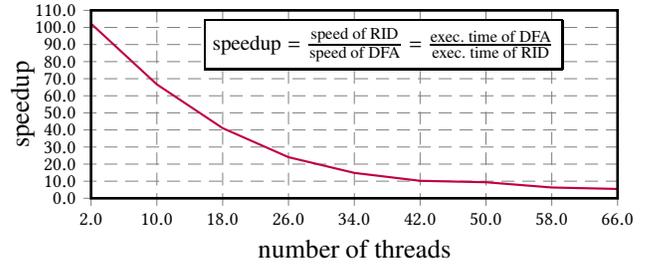
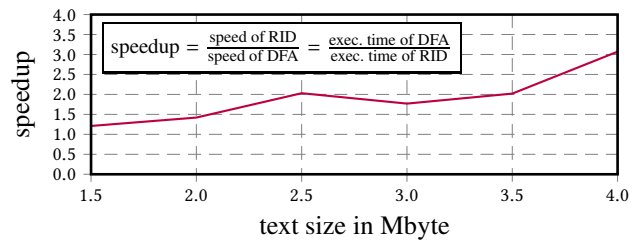
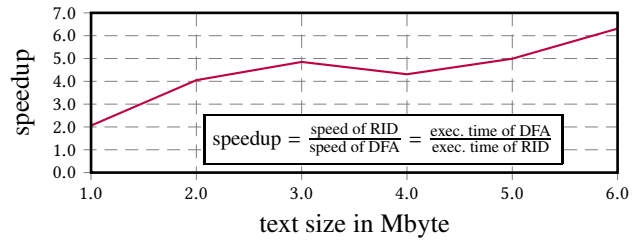

	\begin{subfigure}[c]{0.5\textwidth}
		\begin{center}
			\setpstgraph{gridstyle=dashed, gridwidth=.2pt}
			\rnode{speed1}{\begin{pstgraph}[xmin=2, xmax=66, ymin=0, ymax=10, xgriddiv=8, ygriddiv=10](7,2.5)
				\pstfileplot[linecolor=purple]{files/bible-speedup-vs-thread.data.txt}
				\pstlabel{number of threads}{speedup}
			\end{pstgraph}}
            \fboxsep=0.1cm
            \nput[labelsep=-5.5cm] {-40} {speed1} {\fcolorbox{black}{white}{\footnotesize $\text{speedup} = \frac {\text{speed of RID}} {\text{speed of DFA}} = \frac {\text{exec. time of DFA}} {\text{exec. time of RID}}$}}
		\end{center}
		\caption{Speedup of RID vs DFA for \emph{\bibleHREF} with $4$ Mbyte text.} \label{subfig:bible-speedup-vs-thread}
	\end{subfigure}
	\par
	\vspace{0.5cm}
	\begin{subfigure}[c]{0.5\textwidth}
		\begin{center}
			\setpstgraph{gridstyle=dashed, gridwidth=.2pt}
			\rnode{speed2}{\begin{pstgraph}[xmin=2, xmax=66, ymin=0, ymax=110, xgriddiv=8, ygriddiv=11](7,2.5)
				\pstfileplot[linecolor=purple]{files/regexp-speedup-vs-thread.data.txt}
                \pstlabel{number of threads}{speedup}
			\end{pstgraph}}
            \fboxsep=0.1cm
            \nput[labelsep=-5.5cm] {-40} {speed2} {\fcolorbox{black}{white}{\footnotesize $\text{speedup} = \frac {\text{speed of RID}} {\text{speed of DFA}} = \frac {\text{exec. time of DFA}} {\text{exec. time of RID}}$}}
		\end{center}
		\caption{Speedup of RID vs DFA for \emph{regexp} with $6$ Mbyte text.} \label{subfig:regexp-speedup-vs-thread}
	\end{subfigure}
	\par
	\vspace{0.5cm}
	\begin{subfigure}[c]{0.5\textwidth}
		\begin{center}
			\setpstgraph{gridstyle=dashed, gridwidth=.2pt}
			\rnode{speed3}{\begin{pstgraph}[xmin=1.5, xmax=4.0, ymin=0, ymax=4, xgriddiv=5, ygriddiv=8](7,2.125)
				\pstfileplot[linecolor=purple]{files/bible-speedup-vs-data.data.txt}
				\pstlabel{text size in Mbyte}{speedup}
			\end{pstgraph}}
            \fboxsep=0.1cm
            \nput[labelsep=-6.85cm] {-20} {speed3} {\fcolorbox{black}{white}{\footnotesize $\text{speedup} = \frac {\text{speed of RID}} {\text{speed of DFA}} = \frac {\text{exec. time of DFA}} {\text{exec. time of RID}}$}}
		\end{center}
		\caption{Speedup of RID vs DFA for \emph{\bibleHREF} with $58$ threads.} \label{subfig:bible-speedup-vs-data}
	\end{subfigure}
	\par
	\vspace{0.5cm}
	\begin{subfigure}[c]{0.5\textwidth}
		\begin{center}
			\setpstgraph{gridstyle=dashed, gridwidth=.2pt}
			\rnode{speed4}{\begin{pstgraph}[xmin=1, xmax=6, ymin=0, ymax=7, xgriddiv=5, ygriddiv=7](7,2.125)
				\pstfileplot[linecolor=purple]{files/regexp-speedup-vs-data.data.txt}
                \pstlabel{text size in Mbyte}{speedup}
			\end{pstgraph}}
            \fboxsep=0.1cm
            \nput[labelsep=-5.5cm] {10} {speed4} {\fcolorbox{black}{white}{\footnotesize $\text{speedup} = \frac {\text{speed of RID}} {\text{speed of DFA}} = \frac {\text{exec. time of DFA}} {\text{exec. time of RID}}$}}
		\end{center}
		\caption{Speedup of RID vs DFA for \emph{regexp} with $58$ threads.} \label{subfig:regexp-speedup-vs-data}
	\end{subfigure}
	\caption{\small Speedup for benchmarks \emph{\bibleHREF} and \emph{regexp}.} \label{fig:speedup}
\end{figure}

\noindent
\textbf{\emph{Even and winning benchmarks}.}  The experiment  confirms the  partition (Sect.~\ref{subsec:experimental-speculation}) into the two groups of even and winning benchmarks, as shown in the summary Tab.~\ref{tab:maxspeedup&trans}, where the ratios of transition numbers are reproduced and compared with those of execution times, i.e., speedups (see Fig.~\ref{fig:speedup} for the definition of speedup).
The speedup of RID vs the DFA variant of CSDPA is $> 1$ in three cases and barely $<1$ in two. The RID matches  always the DFA variant and dominates by far the NFA one. In the winning group, the speedup of RID vs the DFA variant is {\red over $3$} in case \emph{\bibleHREF} and {\red over $6$} in \emph{regexp}. The latter  typifies the ideal conditions for top RID performance: a state-minimal NFA such that the equivalent minimal DFA has an exponential blow-up of states.
\par
\begin{table}[tb]
\caption{\small Speedup of RID vs the DFA and NFA variants of CSDPA, i.e., ratio of the execution times of the DFA and NFA variants over RID, with $58$ threads on a shared-memory $64$-core computer. Ratio of the number of transitions of the DFA and NFA variants over RID. All the figures are computed for a text of maximum size.} \label{tab:maxspeedup&trans}
	\begin{center}
		\tabcolsep=0.0cm\def\arraystretch{0.8}
		\begin{tabular}{m{1.25cm}m{0.80cm}wc{1.125cm}wc{1.125cm}wc{0.4cm}wc{1.125cm}wc{1.125cm}wr{1.20cm}}
			&& \multicolumn{2}{c}{\emph{speedup \tiny\rm (RID vs FA)}} && \multicolumn{2}{c}{\emph{transition ratio}} & \emph{text} \\[-0.0cm]
            && \multicolumn{2}{c}{\footnotesize execution time ratio} && \multicolumn{2}{c}{} & \\ \cline{3-4} \cline{6-7} \\[-0.20cm]
			\multicolumn{2}{l}{\emph{benchmark}} & $\frac {\text{DFA}} {\text{RID}}$ & $\frac {\text{NFA}} {\text{RID}}$ && $\frac {\text{DFA}} {\text{RID}}$ & $\frac {\text{NFA}} {\text{RID}}$ & \footnotesize (Mbyte) \\[-0.2cm] \\ \toprule
			\textbf{bigdata} & \tiny even & $1.01$ & $73.24$ && $1.00$ & $1.99$ & $13.10$ \\ \midrule
			\textbf{regexp} & \tiny winning & \red $6.31$ & $56.56$ && $126.99\phantom{00}$ & $14.68\phantom{0}$ & $6.00$ \\ \midrule
			\textbf{\bibleHREF} & \tiny winning & \red $3.07$ & $84.23$ && $8.73$ & $4.19$ & $4.00$ \\ \midrule
			\textbf{\fastaHREF} & \tiny even & $0.94$ & $38.85$ && $1.00$ & $26.26\phantom{0}$ & $0.76$ \\ \midrule
			\textbf{\trafficHREF} & \tiny even & $0.97$ & $109.56\phantom{0}$ && $1.00$ & $1.74$ & $11.20$ \\ \bottomrule
		\end{tabular}
\end{center}
\end{table}
\par\noindent
\textbf{\emph{Sensitivity to chunk number and text length}.} Fig.~\ref{fig:speedup} contains a finer analysis of speedup for the winning group.
The impact of the chunk number (coincident in this experiment with the number of cores) is visible in  Fig.s~\ref{subfig:bible-speedup-vs-thread} and \ref{subfig:regexp-speedup-vs-thread}. The speedup vs DFA decreases when the text of fixed length is cut into more (shorter) chunks, because the overhead of chunk management becomes relevant for short chunks. The opposite effect is shown in Fig.s~\ref{subfig:bible-speedup-vs-data} and \ref{subfig:regexp-speedup-vs-data}, where the speedup vs DFA increases with the text length with a fixed number of chunks.
\par
\subsection{Results for construction time}\label{ssec:constructionTime}
The construction from NFA to RI-DFA is more complex than from  NFA to DFA. On a simple laptop, for the whole \emph{\OndrikHREF} collection of $1.084$ NFAs (see Tab.~\ref{tab:Ondrik}), the former construction takes a time of $2.994$~s (seconds), the latter $146$~s, and their ratio $2.994~\text{s} \, / \, 146~\text{s}$ is about $20$. For  the \emph{\OndrikHREF} collection, the total numbers of states of the given NFAs, and of the constructed DFAs and RI-DFAs, are $2.699.411$, $1.485.483$ and $6.753.792$, respectively.
\par
We argue that this construction cost is much less than what is expected from a worst-case theoretical analysis.
The average number of states of the NFAs of the \emph{\OndrikHREF} collection is $\vert \, Q \, \vert_\emph{avg} = 2.699.411 \, / \, 1.084 = 2.490,2$. In principle, our $\text{NFA} \to \text{RI-DFA}$ construction computes one powerset per each NFA state, while $\text{NFA} \to \text{DFA}$ computes just one, say at a cost $C$. But in reality, each successive powerset computation in $\text{NFA} \to \text{RI-DFA}$ creates fewer and fewer states, so that the overall cost is much less than the product $\vert \, Q \, \vert_\emph{avg} \times C$. As said, the measured time ratio is only $20$, so that the actual individual (i.e., per NFA) cost is $20 \times C$. This  is much lower than the theoretical worst-case cost estimation $2.490 \times C$ obtained by  taking the average NFA size $\vert \, Q \, \vert_\emph{avg}$.
\par
To sum up, although  our construction  computes one powerset per each NFA state,  in reality each powerset execution is just incremental. Thus for the large \emph{\OndrikHREF} collection of big NFAs, the time ratio of $\text{NFA} \to \text{RI-DFA}$ over $\text{NFA} \to \text{DFA}$ is about $20$, i.e., about $1 \, / \, 100$ of the theoretical worst cost. This  is perfectly compatible with a practical use.
\par
We disregard instead the cost of state minimization, which is done only once at the end of both constructions, thus having a negligible overhead.
\section{Conclusion and future work} \label{sec:conclusion}
We have introduced a novel chunk automaton RI-DFA, which can be easily constructed from an NFA. We have proved its correctness, and we have shown that it minimizes the speculation overhead and improves the performance of regular language recognition by means of a data-parallel algorithm CSDPA on multi-core architectures. The main facts follow:
\begin{enumerate}[leftmargin=*]
\item The RID is faster than the  DFA variant of CSDPA for the \emph{winner} benchmarks, and is equally fast  within $10\%$ for the others.
It is always much faster than the NFA variant. \label{enum:mainfact1}
\item The speedup of RID vs the DFA variant grows roughly linearly with the text length in the range considered, provided that enough computing cores are available. \label{enum:mainfact2}
\item The optimal speedup occurs when the  language is such that its NFA is much smaller than its minimal DFA. \label{enum:mainfact3}
\item A large collection of big automata meets statement \eqref{enum:mainfact3}. \label{enum:mainfact4}
\item The cost to construct the RI-DFA is  moderate.
\end{enumerate}
As a consequence  of statements \eqref{enum:mainfact1}, \eqref{enum:mainfact2} and \eqref{enum:mainfact3}, it is not uncommon that for very long texts, or in the case of DFA state explosion, the DFA-based variant of CSDPA fails or takes too long to run to completion, while the RID succeeds.
\par
Although our benchmarks are not representative of any specific application domain, they are not biased to the advantage of the RI-DFA approach. Therefore, the above statements  are very likely to be confirmed by other benchmarks.  Future experimentation will be needed to precisely identify the language patterns that may  mostly benefit from our approach.
\par
Last but not least, the optimization based on the RI-DFA  is compatible with most existing ones, in particular with state-convergence, state speculation exploiting look-back, and even higher order speculation, as described in \cite{Qiu2021ScalableFP}. Therefore, it should be a useful addition for  future parallel finite-state machine implementations.
\par\noindent
\textbf{\emph{Minimality of source automata}.} Of course, an issue for a successful application is the availability of small and possibly minimal NFAs that define the  languages to be recognized. In the following, NFA denotes the  given automaton, $\text{DFA}_\emph{min}$ the minimal equivalent DFA, and $\text{NFA}_\emph{min}$ arbitrarily one of the minimal-state FAs equivalent to NFA (transition minimization is not relevant for speculation reduction). We know from Th.~\ref{th:minimality} that the minimal initial state set is obtained if RI-DFA  is generated from $\text{NFA}_\emph{min}$, which is quite rarely possible in practice.
\par
The simpler languages can be defined by hand as REs (without any guarantee of minimality) and are converted to NFA by standard algorithms, such as GMY~\cite{DBLP:journals/tc/McNaughtonY60}. The result may be much larger than $\text{NFA}_\emph{min}$, yet improvements are possible: using existing more sophisticated $\text{RE} \to \text{NFA}$ converters (see, e.g.,~\cite{DBLP:journals/tcs/GarciaLRA11} and its references), and optimizing the RE prior to its use in a converter (see again, e.g.,~\cite{DBLP:journals/tcs/GarciaLRA11}).
\par
For more complex languages, e.g., those occurring in model checking or formal verification, the NFAs are typically generated by a program, and their level of minimization is often unknown. Unfortunately NFA minimization is  difficult: it is PSPACE-hard \cite{DBLP:conf/focs/MeyerS72}, and the problem cannot even be approximated within a factor of $\mathcal O \, (n)$ unless $\text{P} \equiv \text{PSPACE}$ \cite{DBLP:journals/jcss/GramlichS07}. Many papers have focused on efficient heuristic algorithms for minimization that often, but not always, produce a minimal machine. We refer to some recent papers \cite{DBLP:journals/corr/abs-1301-5585, DBLP:journals/tcs/BianchiniPRR24, DBLP:journals/fuin/LombardyS22, DBLP:journals/acta/BjorklundC21}, where more references are available. It would be  interesting to see how  such heuristics perform on collections of practically relevant NFAs, both in terms of construction time and  approximation to the minimum, and whether such  NFAs pay off in terms of RI-DFA performance. All this is left for future investigation.
\par
At present,  our  initial-state reduction algorithm (Sect.~\ref{subsec:furtherReduction}) is quite capable of  reducing speculation overhead,  and avoids the cost and uncertainty of NFA minimization algorithms.
\section*{Acknowledgments}
To Stefano Zanero and Filippo Carloni for referring us to several RE benchmarks.
\newpage
\bibliography{automatabib}
\end{document}